\documentclass[journal]{IEEEtran}

\ifCLASSOPTIONcompsoc
\usepackage[nocompress]{cite}
\else
\usepackage[noadjust]{cite}
\fi

\usepackage{amsmath,amssymb}
\usepackage{color}

\usepackage{savesym}
\savesymbol{iint}
\usepackage{txfonts}
\restoresymbol{TXF}{iint}
\usepackage{multirow} 
\usepackage{booktabs} 

\usepackage{url}
\usepackage[ruled,lined,boxed,commentsnumbered]{algorithm2e}

\usepackage[utf8x]{inputenc}

\usepackage{tikz}
\usepackage{pgfplots}
\usepgfplotslibrary{fillbetween} 
\usetikzlibrary{arrows,intersections,decorations.pathmorphing,backgrounds,positioning,fit,plotmarks,shadows}
\pgfplotsset{compat=1.10,every axis/.append style={font=\scriptsize}, every legend/.append style={font=\scriptsize}}
\usepackage{color}

\usepackage{fixltx2e}
\usepackage{siunitx}

\newcommand{\EEqref}[1]{\text{Eq.}~\eqref{#1}}

\newcommand{\ETAL}{\textit{et al.},}

\usepackage{amsmath}

\DeclareMathOperator*{\argmax}{arg\,max}

\usepackage{mathrsfs} 

\newcommand{\q}{\ensuremath{\mathbf{q}}}

\newcommand{\uu}{\ensuremath{\mathbf{u}}}

\newcommand{\mathbi}[1]{\ensuremath{\textbf{\textit{#1}}}}
\newcommand{\VV}[1]{\ensuremath{\textbf{\textit{#1}}}}

\newcommand{\Vx}{\mathbi{x}}
\newcommand{\Vy}{\mathbi{y}}

\newcommand{\Vv}{\mathbi{v}}

\newcommand{\BM}[1]{\mathbf{#1}}

\usepackage[bookmarks=true, 
            final=true,
            hyperfootnotes=true,
            ]{hyperref}

\usepackage{graphicx}
\usepackage{grffile} 
\graphicspath{{pics/}{figs/}}

\emergencystretch=\hsize
\newtheorem{proposition}{Proposition}

\usepackage[colorinlistoftodos, textwidth=10mm]{todonotes}
\usepackage{marginnote}

\newcommand{\RCommentLeft}[1]{{%
 \let\marginpar\marginnote
 \reversemarginpar
 \todo[color=blue!20!white]{{#1}}}}

\newcommand{\RCommentLeftCap}[2][]{{%
 \let\marginpar\marginnote
 \reversemarginpar
 \todo[color=blue!20!white, caption={#2}]{{#1}}}}

\renewcommand{\@todonotes@drawMarginNoteWithLine}{%
\begin{tikzpicture}[remember picture, baseline=-0.75ex]%
    \node [coordinate] (inText) {};%
\end{tikzpicture}%
\marginnote[{
    \@todonotes@drawMarginNote%
    \@todonotes@drawLineToLeftMargin%
}]{
    \@todonotes@drawMarginNote%
    \@todonotes@drawLineToRightMargin%
}%
}

\begin{document}

\title{Single- and Multiple-Shell Uniform Sampling Schemes for Diffusion MRI Using Spherical Codes}

\author{Jian~Cheng\IEEEauthorrefmark{1},
        Dinggang~Shen,
        Pew-Thian~Yap\IEEEauthorrefmark{1},
        and~Peter~J.~Basser\IEEEauthorrefmark{1}
\thanks{J.~Cheng and P.-J.~Basser are with SQITS, NIBIB, NICHD, National Institutes of Health. (Emails: jian.cheng@nih.gov, pjbasser@helix.nih.gov)}
\thanks{D.~Shen and P.-T.~Yap are with the Department of Radiology and Biomedical Research Imaging Center (BRIC), the University of North Carolina at Chapel Hill. 
D.~Shen is also with Department of Brain and Cognitive Engineering, Korea University, Seoul 02841, Republic of Korea. 
(Emails: dgshen@med.unc.edu, ptyap@med.unc.edu)}
\thanks{\IEEEauthorrefmark{1} denotes the corresponding authors.}
\thanks{Copyright (c) 2017 IEEE. Personal use of this material is permitted. However, permission to use this material for any other purposes must be obtained from the IEEE by sending a request to pubs-permissions@ieee.org.}
}

\markboth{IEEE Transactions on Medical Imaging, 2017}%
{J. Cheng \MakeLowercase{\textit{et al.}}}
%




\maketitle

\begin{abstract}

In diffusion MRI (dMRI), a good sampling scheme is important for efficient acquisition and robust reconstruction. 
Diffusion weighted signal is normally acquired on single or multiple shells in $\q$-space. 
Signal samples are typically distributed uniformly on different shells to make them invariant to the orientation of structures within tissue, or the laboratory coordinate frame. 
The Electrostatic Energy Minimization (EEM) method, originally proposed for single shell sampling scheme in dMRI, 
was recently generalized to multi-shell schemes, called Generalized EEM (GEEM). 
GEEM has been successfully used in the Human Connectome Project (HCP). 
However, EEM does not \emph{directly} address the goal of optimal sampling, i.e., achieving large angular separation between sampling points.
In this paper, we propose a more natural formulation, called Spherical Code (SC), to directly maximize the minimal angle between different samples in single or multiple shells.  
We consider not only continuous problems to design single or multiple shell sampling schemes, 
but also discrete problems to uniformly extract sub-sampled schemes from an existing single or multiple shell scheme, and to order samples in an existing scheme. 
We propose five algorithms to solve the above problems, 
including an incremental SC (ISC), a sophisticated greedy algorithm called Iterative Maximum Overlap Construction (IMOC), 
an 1-Opt greedy method, 
a Mixed Integer Linear Programming (MILP) method, and a Constrained Non-Linear Optimization (CNLO) method. 
%
To our knowledge, this is the first work to use the SC formulation for single or multiple shell sampling schemes in dMRI. 
Experimental results indicate that 
SC methods obtain larger angular separation and better rotational invariance than the state-of-the-art EEM and GEEM. 
The related codes and a tutorial have been released in DMRITool~\footnote{\label{fn:dmritool}\url{https://diffusionmritool.github.io/tutorial_qspacesampling.html}}.

\end{abstract}

\begin{IEEEkeywords}
  Sampling, q-space, Multiple Shells, Uniform Spherical Sampling, Diffusion MRI, Spherical Codes
\end{IEEEkeywords}

%
\IEEEpeerreviewmaketitle

\section{Introduction}

Diffusion MRI (dMRI) is a unique imaging methodology to explore white matter structure and architecture in the human brain by measuring diffusion of water molecules. 
In dMRI, signal measurements are a limited number of samples of the diffusion signal $E(\q)$ in 3D $\q$-space. 
Reconstruction in dMRI entails fitting the signal samples by using a well-designed model, including Diffusion Tensor Imaging (DTI)~\cite{Basser1994}, 
and various models in High Angular Resolution Diffusion Imaging (HARDI)~\cite{TuchMRM2002,Tuch2004,Cheng_PDF_MICCAI2010,ozarslan_NI13,Descoteaux2007,Cheng_ODF_MICCAI2010,tournier_NI2007,cheng_NI2014,jeurissen_NI2014}. 
After model fitting, meaningful quantities can be estimated, including diffusion tensor and fractional anisotropy in DTI~\cite{Basser1994}, 
the ensemble average propagator, diffusion Orientation Distribution Function (dODF), fiber ODF (fODF), 
and the Generalized Fractional Anisotropy (GFA) in HARDI~\cite{TuchMRM2002,Tuch2004,Cheng_PDF_MICCAI2010,ozarslan_NI13,Descoteaux2007,Cheng_ODF_MICCAI2010,tournier_NI2007,cheng_NI2014}, etc. 
An appropriate data sampling scheme in $\q$-space is important for all such dMRI acquisition and reconstruction applications in order to recover as much information as possible by using a minimal number of measurements. 

The optimal sampling scheme that yields the best reconstruction quality is dependent on the diffusion model, reconstruction algorithm, and the ground truth diffusion signal. 
1) A typical DTI sampling scheme is a single shell scheme with about 30 samples and a low b value ($\leq \SI{1200}{s/mm^2}$)~\cite{jones_MRM99}, 
while multi-shell schemes are preferred (or necessarily needed in some cases) with a higher maximal b value ($\geq \SI{2000}{s/mm^2}$) in HARDI methods~\cite{ozarslan_NI13,Cheng_PDF_MICCAI2010,jeurissen_NI2014,cheng_NI2014}. 
2) Even for the same diffusion model, different reconstruction algorithms, including least squares, robust non-linear estimation with various regularizations, etc., may have different preferred schemes~\cite{cheng:handbook2016}. 
For a given diffusion model, there may be several model fitting (e.g., reconstruction) algorithms. 
Robust estimation may prefer larger b values due to its robustness to noise~\cite{Koay_JMR2006,chang:MRM2005}, 
and compressed sensing based reconstruction could reduce the number of data samples, compared with traditional least square based reconstruction~\cite{cheng_L1SPFI_CDMRI2011,cheng_MICCAI2015}. 
3) Different variabilities of DWI signals may prefer different schemes. 
An efficient scheme should allocate more samples in the domains with larger signal changes than in domains with smaller signal changes. 
Error propagation analysis showed that the optimal b value for DTI is inversely proportional to the mean diffusivity of the latent signal~\cite{jones_MRM99}. 
Therefore, it is infeasible to develop a general optimal sampling scheme that works best for all signal variabilities and all reconstruction methods. 
However, a necessary property for an optimal sampling scheme is that 
the samples should be spherically uniformly distributed with no directional preference, 
because 1) an optimal sampling scheme should be invariant to the orientation of tissue structures and the laboratory coordinate frame in which the data is scanned; 
2) we have to design one single scheme for all different variabilities of signals in human brain. 
Thus, existing works~\cite{jones_MRM99,caruyer_CDMRI11,caruyer_MRM13,zhan:ISBI2011,koay_MP2012,koay_MRM2014,dubois:2006,cook_JMRI2007,desatis:ismrm2011} 
focus on designing single or multiple shell spherically uniform sampling schemes with input parameters $S, \{K_s\}_{s=1}^S$, 
where $S$ is the number of shells and $K_s$ is the number of samples in the $s$-th shell. 
These parameters are user-determined based on the consideration of diffusion signals, models, reconstruction algorithms, noise level, and the scan time limit. 

Uniform single shell sampling schemes are widely used in dMRI, where samples in $\q$-space are uniformly distributed on a sphere (i.e., a fixed b-value). 
Spherical tessellation using spherical polyhedrons can generate uniform single shell schemes with some specific fixed numbers of samples determined by polyhedron types~\cite{TuchMRM2002}. 
On the other hand, the Electrostatic Energy Minimization (EEM) method proposed in dMRI by Jones~\ETAL~\cite{jones_MRM99} is the most popular way to generate a uniform single shell sampling scheme with an arbitrary number of samples. 
EEM considers the samples as electrons on sphere, and estimates the sample configuration by minimizing the electrostatic energy based on Coulomb's law: 
\begin{equation}\label{eq:EEM}
\begin{footnotesize}
  \min_{\{\uu_i\}_{i=1}^K} \sum_{i < j} f_2(\uu_i, \uu_j), \quad \ f_\alpha(\uu_i, \uu_j)= \frac{1}{\|\uu_i-\uu_j\|_2^\alpha} + \frac{1}{\|\uu_i+\uu_j\|_2^\alpha}, 
\end{footnotesize}%
\end{equation}
where the second term in $f_\alpha(\uu_i, \uu_j)$ is used because antipodal symmetric samples provide redundant information in dMRI data reconstruction. 
The problem in~\EEqref{eq:EEM} is the so-called Thomson problem~\footnote{\url{http://en.wikipedia.org/wiki/Thomson_problem}}.
Some best known solutions of EEM have been collected in CAMINO~\cite{cook_ISMRM06}. 
For the multi-shell case, recently it was reported that staggered samples in different shells are better for reconstruction, 
compared with repeated spherical samples in different shells~\cite{caruyer_CDMRI11,caruyer_MRM13,zhan:ISBI2011,koay_MP2012,desatis:ismrm2011}. 
\cite{zhan:ISBI2011} and~\cite{koay_MP2012} tried to design a multi-shell scheme with $S$ shells by separating a single shell scheme into $S$ subsets respectively in different ways to minimize the electrostatic energy. 
\cite{desatis:ismrm2011} proposed to a rotate single shell scheme and use the rotated scheme in different shells. 
\cite{caruyer_CDMRI11} and~\cite{caruyer_MRM13} generalized EEM from the single shell case to the multiple shell case, called Generalized EEM (GEEM), by considering the electrostatic energies both in each individual shell and in the combined shell with all samples. 
The obtained multi-shell schemes by GEEM showed large angular separations in shells than~\cite{zhan:ISBI2011,koay_MP2012}, 
and have been successfully used in the Human Connectome Project (HCP)~\cite{sotiropoulos_HCP_NI13}. 
Although EEM and GEEM are widely used, the electrostatic energy formulation does not directly maximize the angular separation between measurements. 
Samples in $\q$-space are not electrons, and it is still not clear how electrostatic energy relates to dMRI reconstruction. 

A good sampling scheme should have large angular separation such that the reconstruction has large angular resolution and good rotational invariance. 
Thus we propose a mathematically more natural way for sampling scheme design, which directly maximizes the minimal angular difference between sampling points, i.e., \emph{covering radius}, on the unit sphere. 
This way to determine sample configuration on sphere is essentially called 
the Spherical Code (SC) formulation~\footnote{\label{fn:sc}\href{http://mathworld.wolfram.com/SphericalCode.html}{http://mathworld.wolfram.com/SphericalCode.html}}, 
or Tammes problem~\footnote{\url{https://en.wikipedia.org/wiki/Tammes_problem}}. 
In many diffusion MRI applications, the information redundancy is considered in local spatial neighborhoods and local angular neighborhoods~\cite{hagmann:NI2006,stjean:MIA2016,andersson:NI2016,saghafi:HBM2017}, 
which is based on a natural assumption that if the samples $\uu_i$ and $\uu_j$ are close in the sphere, then the DWI signal samples $E(q\uu_i)$ and $E(q\uu_j)$ are also close. 
Thus, based on this assumption, maximizing the separation angles between nearest sampling points is related to minimizing the information redundancy in measured DWI samples. 
The SC formulation directly maximizes the minimal separation angles between the nearest samples. 
While in the EEM formulation, the electrostatic energy is determined by distances between all sample pairs, not just the nearest samples. 
Moreover, the covering radius (i.e., minimal separation angle between samples) is a natural and direct measurement of ``angular resolution'' of spherical sampling schemes, compared with the electrostatic energy. 
Some pioneer works of HARDI uses finer tessellation gradient table (i.e., larger number of samples with smaller separation angles) to investigate the non-Gaussianity of DWI signals~\cite{TuchMRM2002,frank_MRM2002}. 
Wedeen~\ETAL~\cite{Wedeen2005} used $\sqrt{\frac{4\pi}{K}}$ to roughly measure the angular resolution of schemes. 
Caruyer~\ETAL~\cite{caruyer_CDMRI11,caruyer_MRM13} evaluated schemes by different methods using the covering radius. 
Thus, it is natural to directly maximize the covering radius of sampling schemes.

SC is related with the famous \emph{thirteen spheres problem} raised by Isaac Newton, and has been well studied in mathematics~\cite{toth_1949,conway_packing_1996,musin_2015}. 
T\'oth~\cite{toth_1949} proved an upper bound for the covering radius of $K$ samples on a sphere. 
There is also a collection of best known solutions for the SC problem with the antipodal symmetry constraint in $\mathbb{S}^2$~\cite{conway_packing_1996}~\footnote{\label{fn:neil}\href{http://neilsloane.com/grass/dim3/}{http://neilsloane.com/grass/dim3/}}. 
It can be proven that the SC problem is equivalent to EEM problem in~\EEqref{eq:EEM} when $\alpha \to \infty $. 
Thus, Papadakis~\ETAL~\cite{papadakis_MRI00} tried to approximately solve SC problem by iteratively increasing $\alpha$ in~\EEqref{eq:EEM}. 
However as $\alpha$ increases, the optimization problem has more local minima and becomes more unstable 
because the cost function tends to be discontinuous and floating-point overflow occurs~\cite{papadakis_MRI00}. 
Since the SC formulation directly maximizes the minimal distance between samples, it yields larger angular separation than EEM formulation. 
However to our knowledge, the SC formulation and the collected best known solutions have not yet been used in the dMRI domain. 
Although SC is well-studied in the mathematical literature, 
its current formulation is limited to a continuous single shell, and is not applicable to multiple shells, 
nor to the discrete problem to uniformly extract subsets from an existing scheme. 

In this paper, we propose the SC formulation in diffusion MRI with total five algorithms, 
not only for continuous problems to design single and multiple shell schemes, but also for discrete problems to uniformly sub-sample an existing scheme, and for ordering an existing scheme. 
Preliminary parts of this work were reported in two conference papers~\cite{cheng_MICCAI2014,cheng_MICCAI2015_sampling}. 
In this paper, we categorize sampling scheme design problems in dMRI, propose total 5 algorithms in a unified framework of minimizing the mean covering radius, 
and provide additional examples, results, derivations, and insights. 

The paper is organized as follows. 
Section~\ref{sec:sc_sampling} proposes the SC formulation 
and shows how the existing sampling scheme design problems in dMRI can be formulated as optimization problems using SC. 
Section~\ref{sec:alg_sc} proposes 5 algorithms to solve the problems in Section~\ref{sec:sc_sampling}. 
Section~\ref{sec:exp} demonstrates some experimental results of synthetic data and real data.


\section{Spherical Code (SC) and Sampling Problems in DMRI}
\label{sec:sc_sampling}



\subsection{Spherical Code (SC) Formulation}
\label{sec:SC}
 
For a given set of spherical samples $\{\uu_i \in \mathbb{S}^2\}_{i=1}^K$, we define the \emph{covering radius}~\textsuperscript{\ref{fn:sc}} for the $i$-th sample as 
\begin{equation}\label{eq:CoveringRadius_i}
  d_i(\{\uu_l\}_{l=1}^K)= \min_{j \neq i, \forall j } \arccos |\uu_i^T \uu_j |. 
\end{equation}
where the absolute value operator is used because the antipodal symmetric samples have the same role in dMRI data reconstruction. 
We also define the \emph{covering radius} for all samples as the minimal angular distance between all sample pairs, i.e.,
\begin{equation}\label{eq:CoveringRadius}
  d(\{\uu_l\}_{l=1}^K)= \min_i d_i(\{\uu_l\}_{l=1}^K) =  \min_{j\neq i, \forall i, \forall j } \arccos |\uu_i^T \uu_j |, 
\end{equation}
The SC formulation on a single shell is to find $K$ samples  $\{\uu_i\}_{i=1}^K$ 
such that the covering radius is maximized: 
\begin{equation}\label{eq:SC_singleShell}
   \max_{\{\uu_i \in \mathbb{D} \}_{i=1}^K} d(\{\uu_i\}_{i=1}^K)
\end{equation}
where $\mathbb{D} \subseteq \mathbb{S}^2$ is the solution domain. 
If $\mathbb{D}=\mathbb{S}^2$, \EEqref{eq:SC_singleShell} is a continuous SC (CSC) problem for selecting $K$ points from the continuous sphere $\mathbb{S}^2$. 
If $\mathbb{D}=\{\uu_n\}_{n=1}^N$, a set of $N$ predetermined points on $\mathbb{S}^2$, then~\EEqref{eq:SC_singleShell} is a discrete SC (DSC) problem for selecting $K$ from $N$ points. 
For CSC, 
\cite{conway_packing_1996} proposed to iteratively optimize a continuous cost function that approximates the original cost function in~\EEqref{eq:SC_singleShell}. 
The authors also released a collection of best known solutions for the SC problem in $\mathbb{S}^2$~\cite{conway_packing_1996}~\textsuperscript{\ref{fn:neil}}. 
Note that the original SC problem only considers CSC, i.e., $\mathbb{D}=\mathbb{S}^2$, 
while in this paper we propose both CSC and DSC formulations and generalize them to multi-shell case for designing sampling schemes in dMRI.

For multi-shell sampling, SC is to find a set of points $\{\uu_{s,i}\}$ by maximizing the weighted mean of covering radii: 
\begin{equation}\label{eq:SC_multiple}
  \max_{\{\uu_{s,i} \in \mathbb{D} \}} wS^{-1} \sum_{s=1}^{S}  d(\{\uu_{s,i}\}_{i=1}^{K_s})  + (1-w) d(\{\uu_{s,i}\}_{i=1,\dots,K_s; s=1,\dots, S}), 
\end{equation}
where $S$ is the number of shells, $K_s$ is the number of points on the $s$-th shell, $\uu_{s,i}$ is the $i$-th point on the $s$-th shell, 
and $w$ is a weighting factor between the mean covering radius of the $S$ shells and the covering radius of the combined shell containing all points from the $S$ shells. 
$w$ is normally set to $0.5$. 
Because of the second term, the estimated samples in different shell are staggered if $w\neq 1$.

\subsection{Uniform Sampling Scheme Design Problems in dMRI}
\label{sec:problems}

Spherical sampling problems in diffusion MRI can be classified into two categories, 
\textbf{the continuous category (P-C)} which determines samples in the continues sphere, 
and \textbf{the discrete category (P-D)}, i.e., the sub-sampling problem, which determines samples in a discrete set of spherical samples. 
These two problem categories include the following sampling scheme design problems.

\begin{itemize}
  \item 
    \textbf{Single-shell continuous problem (P-C-S)}. 
    Given a number $K$, 
    how does one determine $K$ uniform samples on sphere, such that these samples are separated as far as possible? 
    The problem is for single shell scheme design.
    P-C-S can be solved by CSC in~\EEqref{eq:SC_singleShell} with $\mathbb{D}=\mathbb{S}^2$.
  \item 
    \textbf{Multi-shell continuous problem (P-C-M)}. 
    Given $S$ numbers $\{K_s\}_{s=1}^S$, 
    how does one determine a scheme with $S$ shells, $K_s$ points in the $s$-th shell, 
    such that samples are separated as far as possible not only in each single shell, but also in the combined shell with all samples? 
    The problem is for multi-shell sampling scheme design.
    P-C-M can be solved by CSC in~\EEqref{eq:SC_multiple} with $\mathbb{D}=\mathbb{S}^2$.
  \item 
    \textbf{Single subset from single set problem (P-D-SS)}. 
    Given $N$ known points $\{\uu_i\}_{i=1}^N$ on sphere, 
    how does one uniformly select $K$ samples from the given $N$ samples, 
    such that these samples are separated as far as possible? 
    The problem is to reduce the number of samples in an existing single shell scheme. 
    It also relates to P-C-S, because after discretizing the continuous sphere using many points, solving P-D-SS is an approximation of solving P-C-S.
    P-D-SS can be solved by DSC in~\EEqref{eq:SC_singleShell} with $\mathbb{D}=\{\uu_i\}_{i=1}^N$. 
  \item 
    \textbf{Multiple subsets from multiple sets problem (P-D-MM)}. 
    Given points $\{\uu_{s,i}\}$, $i=1,2,\dots,N_s, s=1,2,\dots,S$, 
    how does one uniformly select $K_s$ from the $N_s$ samples for the $s$-th shell, 
    such that samples are separated as far as possible not only in each single shell, but also in the combined shell with all samples? 
    P-D-MM is to reduce the number of samples in an existing multi-shell scheme.
    It can be solved by DSC in~\EEqref{eq:SC_multiple} with $\mathbb{D}=\{\uu_{s,i}\}$. 
  \item 
    \textbf{Multiple subsets from single set problem (P-D-MS)}. 
    Given a single set of points $\{\uu_i\}_{i=1}^N$ on sphere, 
    how does one uniformly select multiple sets of samples from a given single set of samples, 
    such that the samples are separated as far as possible not only in each set, but also in the combined set with all samples? 
    It relates to P-C-M by discretizing the continuous sphere using many points. 
    P-D-MS can be solved by DSC in~\EEqref{eq:SC_multiple} with $\mathbb{D}=\{\uu_i\}_{i=1}^N$. 
    \cite{zhan:ISBI2011,koay_MP2012} actually tried to solve P-D-MS by minimizing electrostatic energy in different heuristic search ways. 
    We propose to solve it by using a better algorithm called Mixed Integer Linear Programming (MILP).
  \item 
    \textbf{Acquisition ordering problem (P-O)}.
    Given a single or multiple shell scheme, how does one optimize the acquisition order of the samples, 
    such that an interruption at any point of the acquisition can still yield a nearly uniform sampling scheme~\cite{dubois:2006,cook_JMRI2007}? 
    This problem can be solved by using incremental SC (ISC) as described in~\ref{sec:ISC}.
\end{itemize}

P-C-S is a special case of P-C-M, and P-D-SS is a special case of P-D-MM (with $S=1$ and $w=1$). 
In the following context, although P-O belongs to the discrete category, 
we will use P-C to denote P-C-S and P-C-M, P-D to denote P-D-SS, P-D-MM, and P-D-MS.

\section{Algorithms for the SC formulation}
\label{sec:alg_sc}

In this section, we propose five algorithms to solve the above sampling scheme design problems. 
Table~\ref{tab:algorithm_problems} summarizes the algorithms applicable to the corresponding problems. 
Note that although one algorithm may work for several problems, only the problems in bold are suggested for good performance. 
ISC and IMOC do not need an initialization. 
MILP does not require an initialization, although a good initialization may reduce the computation time. 
The algorithms in brackets in the third column are suggested for good initializations. 
The algorithms can also be cascaded, i.e., one algorithm can use the result by another algorithm as the initialization. 
Table~\ref{tab:algorithm_problems_2} lists the suggested algorithms for P-O, P-C and P-D, 
where IMOC + 1-Opt + CNLO means CNLO uses the scheme by IMOC + 1-Opt as its initialization. 

\begin{table}[!t]
  \caption{\label{tab:algorithm_problems}Proposed algorithms to solve the problems listed in Section~\ref{sec:problems}.}
  \begin{center}
  \begin{tabular}{ c | c | c }
    \hline 
    Algorithms & Problems  & Initialization    \\
    \hline 
    ISC  &   P-C, P-D, \textbf{P-O}   & No      \\
    IMOC &   \textbf{P-C}, P-D &  No  \\
    1-Opt  & \textbf{P-C}, P-D & Yes (IMOC) \\
    CNLO &   \textbf{P-C} & Yes (IMOC+1-Opt) \\
    MILP &    P-C, \textbf{P-D} & Not required (IMOC + 1-Opt)\\
    \hline 
  \end{tabular} 
  \end{center}
\end{table}

\begin{table}[!t]
  \caption{\label{tab:algorithm_problems_2}Problems listed in Section~\ref{sec:problems} and suggested algorithms.}
  \begin{center}
  \begin{tabular}{ c | c | c }
    \hline 
    Problems & Suggested algorithms  & Initialization     \\
    \hline 
    P-C &   IMOC + 1-Opt + CNLO &  No  \\
    P-D  &  MILP & Not required (IMOC+1-Opt) \\
    P-O  &   ISC   & No      \\
    \hline 
  \end{tabular}
  \end{center}
\end{table}

\subsection{Incremental SC (ISC)} 
\label{sec:ISC}

The similar incremental strategy used for incremental EEM (IEEM)~\cite{deriche_MIA09} and incremental GEEM (IGEEM)~\cite{caruyer_CDMRI11,caruyer_MRM13} 
can be also used to propose a greedy incremental SC estimation, called incremental SC (ISC). 
ISC works for P-D in both~\EEqref{eq:SC_singleShell} and~\EEqref{eq:SC_multiple} with $\mathbb{D}=\{\uu_i\}_{i=1}^N$.
At step $k$, we estimate one point $\uu\in\mathbb{D}$ that maximizes the cost function based on the $k-1$ points estimated in previous iterations. 
ISC can be also applied to P-C, i.e., 
$\mathbb{D}=\mathbb{S}^2$, by approximating $\mathbb{S}^2$ using a large number of uniformly distributed points. 
In practice, we normally use $20481$ points from a 7 order tessellation of the icosahedron. 
ISC can obtain reasonable schemes in seconds. 
For P-O to optimize acquisition order of a given scheme $\{\uu_{s,i}\}$, 
we can use ISC with $\mathbb{D}=\{\uu_{s,i}\}$. 
In this way, the first partial set of samples are nearly uniformly separated. 
It was reported in~\cite{pierpaoli:2010:artifacts} that optimal ordering reduces the risk of systematic errors due to gradient heating and subject motion.

\subsection{Iterative Maximum Overlap Construction (IMOC)}
\label{sec:IMOC}

Incremental methods (IEEM~\cite{deriche_MIA09}, IGEEM~\cite{caruyer_MRM13} and ISC)
are to obtain a scheme with reasonable uniform coverage when the acquisition is terminated and only a subset of samples are used. 
These incremental methods all have the same limitations. 
1) Because they are developed to work reasonably well for any subset with first several samples, the cost function (i.e., electrostatic energy or covering radius) of the full scheme is not optimal. 
2) When using a finer uniform sample set, the cost function of the obtained sampling scheme may not be improved, which is beyond our expectation. 

\begin{figure*}[t!]
  \begin{center}
\begin{tabular}{c c c}

\begin{tikzpicture}[x=0.23cm,y=0.23cm]
  
  \node (po) at (0,0) {};
  \draw[draw,thick,color=black] (po) circle (10);

  \node (px) at (0,10) {};
  \draw (px) node[above left] {$\mathbf{u}_1$};
  \draw[thick,color=red,fill] (px) circle (0.3);
  \draw[color=black,dashed] (px) circle (3);
  \draw [yellow,thick,domain=72.746:107.25] plot ({10*cos(\x)}, {10*sin(\x)});
  \node (pnx) at (0,-10) {};
  \draw[color=red] (pnx) circle (0.3);
  \draw[color=black,dashed] (pnx) circle (3);
  \draw [yellow,thick,domain=-72.746:-107.25] plot ({10*cos(\x)}, {10*sin(\x)});
  \draw[densely dashed] (px) -- (pnx);
  
  \node (px) at (2.9661,9.55) {};
  \node (px1) at (-2.9661,9.55) {};
  \draw (0,2) arc (90:72.746:2);
  \draw (-0.2,2) node[above right] {$\theta$};

  \node (px) at (2.9661,9.55) {};
  \draw (px) node[above left] {$\mathbf{u}_2$};
  \draw[thick,color=red,fill] (px) circle (0.3);
  \draw[color=black,dashed] (px) circle (3);
  \draw [yellow,thick,domain=55.492:72.746] plot ({10*cos(\x)}, {10*sin(\x)});
  \node (pnx) at (-2.9661,-9.55) {};
  \draw[color=red] (pnx) circle (0.3);
  \draw[color=black,dashed] (pnx) circle (3);
  \draw [yellow,thick,domain=-72.746:-124.51] plot ({10*cos(\x)}, {10*sin(\x)});
  \draw[densely dashed] (px) -- (pnx);
  
  \node (px) at (5.6652,8.2405) {};
  \draw (px) node[above right] {$\mathbf{u}_3$};
  \draw[thick,color=red,fill] (px) circle (0.3);
  \draw[color=black,dashed] (px) circle (3);
  \draw [yellow,thick,domain=55.492:38.238] plot ({10*cos(\x)}, {10*sin(\x)});
  \node (pnx) at (-5.6652,-8.2405) {};
  \draw[color=red] (pnx) circle (0.3);
  \draw[color=black,dashed] (pnx) circle (3);
  \draw [yellow,thick,domain=-141.76:-124.51] plot ({10*cos(\x)}, {10*sin(\x)});
  \draw[densely dashed] (px) -- (pnx);

  \draw (6,-13) node[above] {$\mathbb{S}^1$};

\end{tikzpicture}
  &
\begin{tikzpicture}
\begin{axis}[
width=0.4\textwidth, 
axis lines=left,
xlabel={$\theta_s$},
ylabel={$\theta_0$},
xtick=\empty,
ytick=\empty,
xmin=0, xmax=1.2,
ymin=0, ymax=2.4
]

\addplot[color=black,mark=square*,nodes near coords,only marks,
point meta=explicit symbolic]
table[meta=label] {
x y label
0.95 0.5342  $\theta_s=\theta_s^{\text{ub}}$
0.2 1.9592  $\theta_0=\theta_0^{\text{ub}}$
0.682307 1.4071 \text{IMOC solution}
 };

\addplot[black,name path=A1,domain=0:0.21] {1.9592}; 
\addplot[thick,red,name path=A2,domain=0.2:0.95] {-exp(x^2)+3}; 
\addplot[name path=B1,domain=0:0.21] {0}; 
\addplot[name path=B2,domain=0.2:0.95] {0}; 
\addplot[gray!40] fill between[of=A1 and B1];
\addplot[gray!40] fill between[of=A2 and B2];

\addplot[const plot]
coordinates
{(0.95,0)
 (0.95,0.5342)
 };

\addplot[dashed,thick,blue,domain=0:0.95] {2.0623*x}; 
\addplot[dashed,black,domain=0.21:0.95] {1.9592}; 
\addplot[dashed,const plot]
coordinates
{(0.95,0.5342)
 (0.95,1.9592)
 };

\end{axis}
\end{tikzpicture}
&
\includegraphics[width=0.3\textwidth]{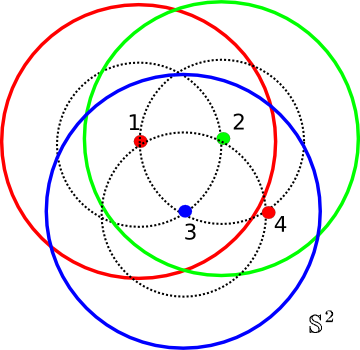} \\
(a) & (b) & (c)

\end{tabular}
  \end{center}
  \caption{\small\label{fig:IMOC_search}
  \textbf{Sketch map of IMOC}. 
(a) demonstrates MOC algorithm~\ref{alg:MOC_single} for P-C-S in $\mathbb{S}^1$, 
where $\uu_1$, $\uu_2$ and $\uu_3$ are selected one by one with a given $\theta$, and the yellow arc denotes the total coverage set.  
Figures (b) and (c) are sketch maps for the multi-shell case P-C-M. 
(b) shows the compromise between the covering radius $\theta_0$ in the combined shell and the covering radius $\theta_s$ in the $s$-th shell, 
and demonstrates the binary search path (the dashed blue line segment) and the obtained covering radii in IMOC. 
The grey area contains all feasible covering radii, and the red curve denotes all Pareto optimal covering radii. 
(c) demonstrates MOC for 3 shell case. 
Three colors denote three shells. 
The numbers near the points are the selection orders in MOC for the points. 
Point 1 (i.e., $\uu_{1,1}$) is first selected for shell 1, then point 2 ($\uu_{2,1}$) for shell 2 and point 3 ($\uu_{3,1}$) for shell 3. 
The red circle around $\uu_{1,1}$ denotes its coverage set $C(\uu_{1,1},\theta_1)$ which is added to $\text{CS}_{1}$, 
and the dashed black circle around $\uu_{1,1}$ denotes $C(\uu_{1,1},\theta_0)$ which is added to $\text{CS}_{0}$. 
Point 4 ($\uu_{1,2}$) is selected based on $\text{CS}_s$, $\forall s=0,1,2,3$.
} 
\end{figure*}
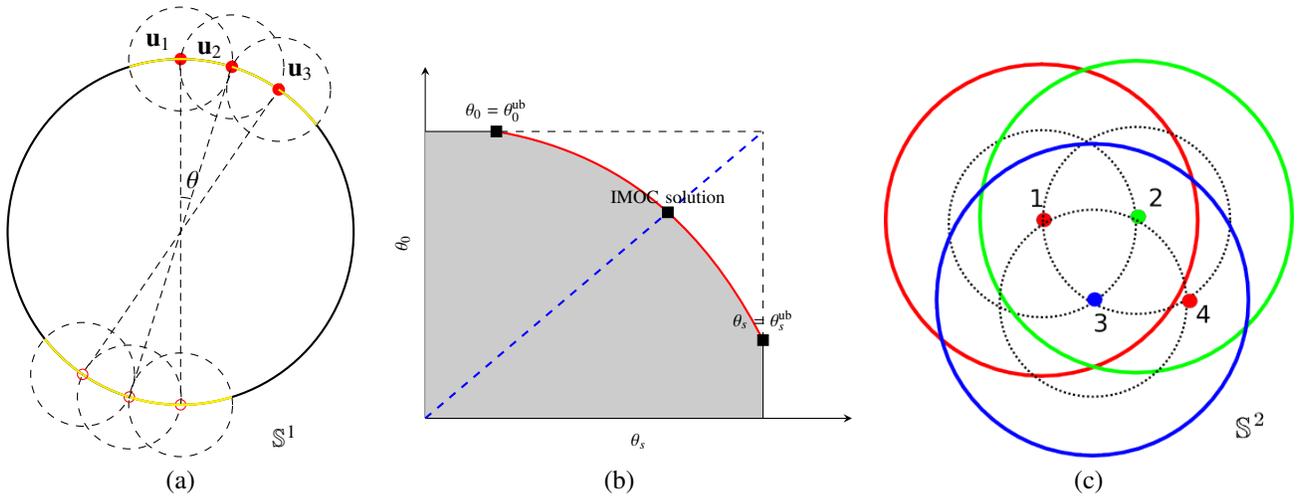

{\footnotesize
\begin{algorithm}[t!]
\caption{\label{alg:IMOC_single}\textbf{Iterative Maximum Overlap Construction (IMOC) for Single shell Scheme Design:}}
\SetAlgoLined
  \KwIn{$K$, $\mathbb{D}$ }
\KwOut{$K$ samples $\{\uu_{i}\}_{i=1}^{K}$.}
// binary search $\theta$ from $(0,\theta^{\text{ub}})$. $\theta^{\text{ub}}$ is the upper bound of the covering radius \;
  $\theta^0=0$, $\theta^1=\theta^{\text{ub}}$ \;
  \Repeat{$\theta$ does not change}
{
  $\theta = (\theta^0+\theta^1)/2$ \;
  [IsSatisfied, $\{\uu_i\}$] = \textbf{MOC}($\theta$, $K$, $\mathbb{D}$)\;
    \lIf{IsSatisfied}{$\theta^0=\theta$}; \lElse{$\theta^1=\theta$}\;
}
\end{algorithm}
}

{\footnotesize
\begin{algorithm}[t!]
\caption{\label{alg:MOC_single}\textbf{Maximum Overlap Construction (MOC) for Single shell Scheme Design:}}
\SetAlgoLined
  \KwIn{$\theta$, $K$, $\mathbb{D}$ }
\KwOut{IsSatisfied, $\{\uu_{i}\}_{i=1}^{K}$.}
  // Define the coverage of $\Vx$ as $C(\Vx,\theta)= \{\Vy \ | \ \arccos(|\Vy^T\Vx|) < \theta, \Vy\in \mathbb{S}^2 \}$ \;
  // Define $\mathbb{D}-\text{CS}=\{\Vy \ | \ \Vy\in \mathbb{D}, \Vy \notin \text{CS} \}$ \;
Initialize coverage set $\text{CS}$ as an empty set\;
\For{$i\leftarrow 1$ \KwTo $K$}
{
  \lIf{$(\mathbb{D}-\text{CS})$ is empty  }{IsSatisfied = \textbf{False}; \Return }\; 
  \lIf{$i== 1$  }{choose any $\uu_1$}\; 
  \lElse{choose $\uu_i$ in $(\mathbb{D}-\text{CS})$ such that the set $C(\uu_i,\theta) \cap \text{CS}$ has the largest area}\;
  $\text{CS} \leftarrow \text{CS} \cup C(\uu_i,\theta)$\;
}
IsSatisfied=\textbf{True}; \Return;
\end{algorithm}
}

\begin{algorithm}[t!]
\caption{\label{alg:IMOC_multiple}\textbf{IMOC for Multiple shell Scheme Design:}}
\SetAlgoLined
  \KwIn{number of samples for $S$ shells: $\{K_s\}_{s=1}^S$, and $\mathbb{D}$ }
\KwOut{$\{\uu_{s,i}\}_{i=1}^{K_s}$.}
// binary search $\{\theta_s\}$ from $\{(0,\theta_s^{\text{ub}})\}$\; 
  $\theta_s^0=0$, $\theta_s^1=\theta_s^{\text{ub}}$, $\forall s=0,1,\dots,S$ \;
\Repeat{$\theta_s$ does not change, $\forall s=0,1,\dots,S$}
{
  $\theta_s = (\theta_s^0+\theta_s^1)/2$, $\forall s=0,1,\dots,S$ \;
  [IsSatisfied, $\{\uu_{s,i}\}$] = \textbf{MOC}($\{\theta_s\}_{s=0}^S$, $\{K_s\}_{s=1}^S$, $\mathbb{D}$)\;
    \lIf{IsSatisfied}{
$\theta_s^0=\theta_s$, $\forall s=0,1,\dots,S$
    }\; 
    \lElse{
$\theta_s^1=\theta_s$, $\forall s=0,1,\dots,S$
}\;
}
\end{algorithm}

\begin{algorithm}[t!]
\caption{\label{alg:MOC_multiple}\textbf{MOC for Multiple shell Scheme Design:}}
\SetAlgoLined
  \KwIn{$\{\theta_s\}_{s=0}^S$, $\{K_s\}_{s=1}^S$, $\mathbb{D}$ }
\KwOut{IsSatisfied, $\{\uu_{s,i}\}_{i=1}^{K_s}$.}
Initialize coverage sets $\{\text{CS}_s\}_{s=0}^S$ as $S+1$ empty sets, and initialize $N_s =0$, $\forall s\in[1,S]$\;
\For{$n = 1$ \KwTo $\sum_{s=1}^SK_s$}
{
  \lIf{$n==1$}{choose any point as $\uu_{1,1}$, $s\leftarrow 1$, $i\leftarrow 1$}\; 
  \lIf{$1 < n \leq S$}{$s\leftarrow n$, $i\leftarrow 1$, choose $\uu_{s,i}$ in $(\mathbb{D}-\text{CS}_0)$ such that the set $C(\uu_{s,i},\theta_0) \cap \text{CS}_0$ has the largest area }\; 
  \If{$n>S$}
  {
    Set $V$ as an empty set\;
  \For{$s' = 1$ \KwTo $S$}
  {
    \If{$N_{s'}<K_{s'}$ \textbf{and} $(\mathbb{D} - (\text{CS}_{s'} \cup \text{CS}_0))$ is not empty}
    {    
      choose $\Vv_{s'}$ in $(\mathbb{D}-(\text{CS}_{s'} \cup \text{CS}_0))$ such that the overlap set $C(\Vv_{s'},\theta_{s'}) \cap (\text{CS}_{s'} \cup \text{CS}_0 )$ has the largest area denoted as $A_{s'}$ \;
      $V \leftarrow V \cup \{\Vv_{s'}\}$\;
    }
  }
  \eIf{$V$ is empty}{IsSatisfied = \textbf{False}; \Return }
  {
  choose $s$ and $\Vv_s\in V$ such that their corresponding area $A_s$ is the largest one among $\{A_{s'}\}_{s'=1}^S$\; 
  $i \leftarrow N_{s}+1$,\  $\uu_{s,i}\leftarrow \Vv_s$\;
  }
}
  $\text{CS}_s \leftarrow \text{CS}_s \cup C(\uu_{s,i},\theta_s) $;\ \  
  $\text{CS}_0 \leftarrow \text{CS}_0 \cup C(\uu_{s,i},\theta_0) $;\ \
  $N_s\leftarrow N_s+1$\;
}
IsSatisfied=\textbf{True}; \Return;
\end{algorithm}

We propose a sophisticated greedy algorithm, called Iterative Maximum Overlap Construction (IMOC), 
to iteratively search the covering radius and sequentially generate samples with maximal spherical overlap on the sphere. 
IMOC works for both P-C and P-D. 
See the IMOC in Alg.~\ref{alg:IMOC_single} for designing single shell schemes. 
IMOC uses MOC in Alg.~\ref{alg:MOC_single} to verify whether a candidate covering radius $\theta$ is able to construct an acceptable scheme with $K$ samples. 
We define the coverage set of a point $\Vx$ as 
$C(\Vx,\theta)= \{\Vy \mid \arccos(|\Vy^T\Vx|) < \theta, \Vy\in\mathbb{S}^2 \}$. 
With a given candidate $\theta$, MOC generates samples one by one. 
In the $i$-th step, MOC finds the best point $\uu_i$ whose coverage $C(\uu_i,\theta)$ has the largest overlap with the existing total coverage set $\text{CS}$, 
and then adds $C(\uu_i,\theta)$ into $\text{CS}$. 
See Alg.~\ref{alg:MOC_single}. 
If all $K$ points can be obtained by MOC using $\theta$, then the best covering radius is no less than $\theta$, which means $\theta$ can be increased.  
If MOC cannot find $K$ points using $\theta$, then $\theta$ can be decreased. 
Therefore, IMOC performs a binary search of the covering radius $\theta$ from $(0, \theta^{\text{ub}}(2K))$, 
where $\theta^{\text{ub}}(2K)$ in~\EEqref{eq:ub_2K} 
is an upper bound of the covering radius with $K$ points with antipodal symmetry consideration (i.e., total $2K$ points)~\cite{toth_1949}. 
\begin{equation}\label{eq:ub_2K}
  \theta^{\text{ub}}(2K)=\arccos\sqrt{4- \csc^2\left( \frac{\pi K}{6(K-1)} \right) }
\end{equation}
See Fig.~\ref{fig:IMOC_search} (a) for a demonstration of MOC when solving P-C-S in $\mathbb{S}^1$ in the 2D case. 
It is obvious that IMOC yields the \emph{globally optimal solution} in $\mathbb{S}^1$, while ISC and IGEEM cannot. 

IMOC and MOC algorithms for multiple shell sampling scheme design are shown in Alg.~\ref{alg:IMOC_multiple} and~\ref{alg:MOC_multiple}. 
For multiple shell case, IMOC needs to search $(\theta_0, \theta_1,\dots,\theta_S)$ in $S+1$ dimensional space, 
where $\theta_0$ is the covering radius of the combined shell with all samples and $\theta_s$ ($s>0$) is the covering radius of the $s$-th shell. 
There is a compromise between $\theta_0$ and $\theta_s$ ($s>0$). 
The cost function in~\EEqref{eq:SC_multiple} essentially converts a \emph{multi-objective optimization problem} into a single objective optimization using a weighted summation. 
There are infinite \emph{Pareto optimal} covering radii to the multi-objective optimization problem. 
IMOC only searches the covering radii in one dimensional space determined by the upper bounds $\{\theta_s^{\text{ub}}\}_{s=0}^S$, 
which in practice provides good covering radii for both individual shells and the combined shell. 
See  Alg.~\ref{alg:IMOC_multiple} and Fig.~\ref{fig:IMOC_search} (b) and (c). 
MOC in Alg.~\ref{alg:MOC_multiple} first selects one point $\uu_{s,1}$ for each shell. 
Then it selects the best shell $s$ and the best point $\uu_{s,i}$ whose coverage set $C(\uu_{s,i},\theta_s)$ 
has the largest overlap area with the existing total coverage set $\text{CS}_s \cup \text{CS}_0$ among all shells and all candidate points outside the existing total coverage set.

Similarly with ISC, for P-C, IMOC can analytically select the best point in each iteration in 2D case in $\mathbb{S}^1$. 
See Fig.~\ref{fig:IMOC_search} (a). 
However, in $\mathbb{S}^2$, IMOC requires a fine uniform discretization of the sphere to find the best point in each step with the largest overlap area by counting the number of points in overlaping sets. 
For efficient implementation of IMOC, we consider 3 issues. 
1) KD-tree is used for efficient nearest neighbor search. 
2) The overlap area $A_s(\Vx)$ for a candidate point $\Vx$ in the $s$-th shell can be stored for next step. 
In the $i+1$-th step, $A_s(\Vx)$ is different from the $i$-th step and needs to be re-calculated, 
only if the determined point $\uu_{s',i}$ in the $i$-th step satisfies $\arccos(|\Vx^T\uu_{s',i}|)<\theta_0$, if $s\neq s'$, 
or $\arccos(|\Vx^T\uu_{s',i}|)<\theta_s$, if $s= s'$. 
3) In MOC, for P-C, the candidate points are selected from a small \emph{``outside surface''} of $\text{CS}$, 
i.e., $\{\Vx \ | \ \Vx \notin \text{CS}, \exists \Vy\in \text{CS}, \text{s.t.} \arccos(|\Vx^T\Vy|)\leq\delta \}$, 
instead of the entire complementary set $(\mathbb{S}^2-\text{CS})$. 
$\delta$ is set as twice of the covering radius of the fine uniform sampling set used in IMOC. 
It is obvious that in the 2D case, this modification does not change the result because the point in each step of MOC is always in the \emph{``outside surface''}. 
See Fig.~\ref{fig:IMOC_search} (a). 
For P-C-S and P-C-M in $\mathbb{S}^2$, the results by IMOC do not change neither in our experiments, although we cannot give a proof. 
These three issues significantly speed up MOC, 
because the first one reduces the time for each neighborhood search, 
and the other two largely reduce the number of neighborhood searches. 
In our implementation, IMOC only requires a few seconds on an ordinary laptop.


\subsection{1-Opt Algorithm} 
\label{sec:1Opt}

We propose a greedy algorithm called 1-Opt. 
For both P-C and P-D, 
with a given initialization scheme, 
the 1-Opt algorithm at each step finds the best point pair ($\uu_{s,i}$ from current result and $\Vx\in \mathbb{D}$) 
such that after setting $\uu_{s,i}$ as $\Vx$ the covering radius for $\uu_{s,i}$ can be best increased. 
For single shell case, 1-Opt is to increase $d_i$ defined in~\EEqref{eq:CoveringRadius_i}. 
For multi-shell case, 1-Opt is to increase both covering radii of $\uu_{s,i}$ in the combined shell, 
i.e., $d_{s,i}=\argmax_{(s,i)\neq (s',j), \forall j, \forall s'} \arccos |\uu_{s,i}^T\uu_{s',j'}| $ and in the $s$-th shell, i.e., 
$d_i=\argmax_{j\neq i, \forall j} \arccos |\uu_{s,i}^T\uu_{s,j}|$. 
See Alg.~\ref{alg:1opt}. 
1-Opt is to increase the covering radius for every individual sample. 
1-Opt requires an initialization, and a good initialization normally results in a good result. 
In practice, we use the result by IMOC as the initialization of 1-Opt. 

Note that if IMOC is used to design a scheme with many samples for P-C, 
then there sometimes is a hole area in the result of IMOC. 
This is because of the discretization used in IMOC for P-C. 
By applying 1-Opt to the IMOC result, 
even though 1-Opt sometimes cannot increase the overall covering radius for the whole scheme, it can better arrange the points to fix the hole. 
See Fig.~\ref{fig:visual}.

{\footnotesize
\begin{algorithm}[t!]
\caption{\label{alg:1opt}\textbf{1-Opt for Multi-Shell Scheme Design:}}
\SetAlgoLined
  \KwIn{Initialization $\{\uu_{s,i}^0\}$, and domain $\mathbb{D}$ }
  \KwOut{$\{\uu_{s,i}\}$.}
  $\uu_{s,i}=\uu_{s,i}^0$, $\forall s, i$  \;
  \Repeat{Cannot find such point pair}
{
  1) Find a point pair, where one point $\uu_{s,i} \in \{\uu_{s',i'}\} $ and one point $\Vx \in \mathbb{D}$, 
  such that after setting $\uu_{s,i}$ as $\Vx$, $d_{s,i}$ and $d_{i}$ both increase (or one increases, one does not decrease), 
  and there is no other pair can obtain both larger $d_{s,i}$ and larger $d_{i}$\;
  2) Set $\uu_{s,i}$ to $\Vx$, and remove $\Vx$ from $\mathbb{D}$;
}
\end{algorithm}
}

\subsection{Constrained Non-Linear Optimization (CNLO)}
\label{sec:CNLO}

For P-C, SC problem in~\EEqref{eq:SC_singleShell} can be converted into a continuous optimization, 
called Constrained Non-Linear Optimization (CNLO): 
\begin{subequations}\label{eq:CNLO_single}
  \begin{align}
    & \max_{\theta, \{\uu_i\}_{i=1}^K}   \theta  \label{eq:CNLO_single_cost} \\
    \text{s.t.}\ & \arccos(|\uu_i^T\uu_j|) \geq \theta,\ ~\forall i<j\leq K; \label{eq:CNLO_single_c1} \\
  & \uu_i^T\uu_i = 1, \ \forall i; \label{eq:CNLO_single_c2}
\end{align}
\end{subequations}
where~\EEqref{eq:CNLO_single_c1} means all separation angles are larger than $\theta$, 
and~\EEqref{eq:CNLO_single_c2} means the estimated samples are on the unit sphere. 
After the problem is solved, the solution $\theta^*$ is the estimated covering radius and there exists $i$ and $j$ such that $\arccos|(\uu_i^*)^T\uu_j^*|=\theta^*$. 
Note that both cost function and constraints are continuous. 

CNLO in~\EEqref{eq:CNLO_single} has total $\frac{K(K-1)}{2}$ inequality constraints. 
Thus for P-C-S or P-C-M with a large number of samples, CNLO may be very slow due to large number of constraints. 
CNLO is a local optimization method which requires an initialization. 
Thus for a given initialization $\{\VV{p}_i\}$ we propose a sped up version of CNLO to reduce the number of inequality constraints via
\begin{subequations}\label{eq:CNLO_single_local}
  \begin{align}
  & \max_{\theta, \{\uu_i\}_{i=1}^K}   \theta \label{eq:CNLO_single_local_cost} \\
    \text{s.t.}\ & \arccos(|\uu_i^T\uu_j|) \geq \theta,\ \text{if}\ |\VV{p}_i^T\VV{p}_j| \geq \cos(2\delta_0 + \theta^{\text{ub}}),   ~\forall i<j\leq K; \label{eq:CNLO_single_local_c1} \\
    & \arccos(|\uu_i^T\VV{p}_i|) \leq \delta_0, \ \forall i;  \\
  & \uu_i^T\uu_i = 1, \ \forall i; \label{eq:CNLO_single_local_c2}
\end{align}
\end{subequations}
where $\delta_0$ controls the difference between the final results $\{\uu_i\}$ and the initialization $\{\VV{p}_i\}$, 
and the inequality constraints which have two initialization points far from $2\delta_0 + \theta^{\text{ub}}$ do not need to be considered in~\EEqref{eq:CNLO_single_local_c1}, 
because $\theta^{\text{ub}}$ is an upper bound. 
If $\delta_0=0$, then the result is just the initialization. 
If $\delta_0=\pi/2$, then the result of~\EEqref{eq:CNLO_single_local} is the same as the result of~\EEqref{eq:CNLO_single}.  
When setting $\delta_0$ as a small float number, the number of constraints can be largely reduced and CNLO can be sped up.

Similarly, with a given initialization $\{\VV{p}_{s,i}\}$, 
P-C-M in~\EEqref{eq:SC_multiple} can be solved using CNLO in~\EEqref{eq:CNLO_multiple}:
\begin{subequations}\label{eq:CNLO_multiple}
  \begin{align}
    & \max_{\{\theta_s\}, \theta_0, \{\uu_{s,i}\}}  w\frac{1}{S} \sum_{i=1}^S \theta_s + (1-w)\theta_0  \label{eq:CNLO_multiple_cost}\\
    \text{s.t.}\ & \arccos(|\uu_{s,i}^T\uu_{s,j}|) \geq \theta_s,\ \text{if}\ |\VV{p}_{s,i}^T\VV{p}_{s,j}| \geq \cos(2\delta_0 + \theta_s^{\text{ub}}), \nonumber \\
    & \qquad\qquad\qquad\qquad\qquad  ~\forall s,\ \forall i<j\leq K_s; \label{eq:CNLO_multiple_c1} \\
    & \arccos(|\uu_{s,i}^T\uu_{s',j}|) \geq \theta_0, \ \text{if}\ |\VV{p}_{s,i}^T\VV{p}_{s',j}| \geq \cos(2\delta_0 + \theta_0^{\text{ub}}), \nonumber \\ 
    & \qquad\qquad\qquad\qquad  ~\forall s <s',\ ~\forall i\leq K_s,\ ~\forall j\leq K_{s'}; \label{eq:CNLO_multiple_c2}\\
    & \arccos(|\uu_{s,i}^T\VV{p}_{s,i}|) \leq \delta_0, \ \forall s, \forall i;  \\
     & \theta_s \geq \theta_0,\ ~\forall s; \label{eq:CNLO_multiple_c3} \\
     & \uu_{s,i}^T\uu_{s,i} = 1, \ \forall s, i; \label{eq:CNLO_multiple_c4}
\end{align}
\end{subequations}%
$\delta_0$ controls the difference between the result and the initialization. 
\EEqref{eq:CNLO_multiple_c1} means that the separation angles of samples in the $s$-th shell are larger than its corresponding covering radius $\theta_s$. 
\EEqref{eq:CNLO_multiple_c2} means that the separation angles of samples in two different shells are larger than the covering radius $\theta_0$ for the combined shell with all samples. 
\EEqref{eq:CNLO_multiple_c3} means that the covering radii for all single shells are always larger than the covering radius for the combined shell, 
in other words, \EEqref{eq:CNLO_multiple_c3} means that the separation angles of samples in the same shell are larger than $\theta_0$.  

CNLO in~\EEqref{eq:CNLO_single_local} and~\EEqref{eq:CNLO_multiple} are continuous non-convex constrained optimization problems. 
We use sequential quadratic programming (SQP) to solve CNLO. 
In each step, SQP solves a quadratic programming problem which locally approximates the original non-linear problem. 
In practice, we use the SQP solver implemented in NLOPT library~\cite{NLPOT} (\texttt{NLOPT\_LD\_SLSQP}). 
CNLO yields a locally optimal solution with a given initialization.  
We suggest to use the result by IMOC + 1-Opt as the initialization. 
In practice, we set $\delta_0=0.1$, and keep performing CNLO by updating the initialization $\{\VV{p}_i\}$ (or $\{\VV{p}_{s,i}\}$) using the previously obtained result. 

\subsection{Mixed Integer Linear Programming (MILP)}
\label{sec:MILP}

We propose a Mixed Integer Linear Programming (MILP) algorithm to solve P-D. 
For $\mathbb{D}=\{\uu_n\}_{n=1}^N$, ~\EEqref{eq:SC_singleShell} can be solved using MILP:
\begin{subequations}\label{eq:milp_single}
  \begin{align}
  & \max_{\theta, \{h_i\}_{i=1}^N}   \theta \\
  \text{s.t.}\ & \arccos(|\uu_i^T\uu_j|) \geq \theta - (2-h_i-h_j)M,\ ~\forall i>j \label{eq:milp_single_c1} \\
  &  \theta^{\text{lb}} \leq \theta \leq \theta^{\text{ub}}(2K)  \\
  & \sum_{i=1}^N h_i = K; \qquad~h_i=0,1, \ \forall i
\end{align}
\end{subequations}
where $h_i=1$ means that $\uu_i$ is selected as one of the $K$ points, 
$\theta^{\text{lb}}$ is a lower bound of the covering radius, 
which can be set to $0$ or the covering radius from an existing sampling scheme, 
$\theta^{\text{ub}}(2K)$ is an upper bound of the covering radius in~\EEqref{eq:ub_2K}, 
and $M$ is the difference between the maximal distance (denoted as $d_{\text{max}}$) and the minimal distance (denoted as $d_{\text{min}}$) of any two points $\uu_i,\uu_j\in\mathbb{D}$, $i \neq j$. 
After solving MILP, the solution of $\theta$, denoted as $\theta^*$, is the covering radius of the selected $K$ samples. 
The constraint in~\EEqref{eq:milp_single_c1} only takes effect when two points are both selected, i.e., $h_i=h_j=1$, 
and is automatically satisfied when $h_i=0$ or $h_j=0$, 
because we set $M$ large enough such that $\arccos(|\uu_i^T\uu_j|) \geq d_{\text{min}} \geq \theta^* - (2-h_i-h_j)M$, when $h_i+h_j \leq 1$.

Similarly, ~\EEqref{eq:SC_multiple} with $\mathbb{D}=\{\uu_{s,i}\}$ can be solved using MILP:
  \begin{subequations}\label{eq:milp_multi}
\begin{footnotesize}
  \begin{align}
    & \max_{\{\theta_s\}, \{h_{s,i}\} }   w S^{-1} \sum_{s=1}^S \theta_s + (1-w)\theta_0 \label{eq:milp_multi_cost}\\
  \text{s.t.}\ 
  & \arccos(|\uu_{i}^T\uu_{j}|) \geq \theta_s - (2-h_{s,i}-h_{s,j})M,\ ~\forall s, i>j  \label{eq:milp_multi_c1}\\
  & \arccos(|\uu_{i}^T\uu_{j}|) \geq \theta_0 - (2-h_{s,i}-h_{s',j})M,\ ~\forall s, s', i>j \label{eq:milp_multi_c2}\\
  & \theta^{\text{lb}}_s \leq \theta_s \leq \theta^{\text{ub}}(2K_s),~\forall s; \quad~ \theta^{\text{lb}}_0 \leq \theta_0 \leq \theta^{\text{ub}}(2\sum_{s=1}^S K_s) \label{eq:milp_multi_c3}\\
    & \sum_{i=1}^N h_{s,i} = K_s,~\forall s; \quad~\sum_{s=1}^S h_{s,i} \leq 1, \ \forall i; \quad~h_{s,i}=0,1, \ \forall i,s \label{eq:milp_multi_c4}
\end{align}
\end{footnotesize}%
\end{subequations}
where $h_{s,i}=1$ means that $\uu_{i}$ is selected as one of the $K_s$ points on the $s$-th shell, 
$\theta^{\text{lb}}_s$ and $\theta^{\text{ub}}(2K_s)$ are the lower and upper bounds of the covering radius on the $s$-th shell, 
and $\theta^{\text{lb}}_0$ and $\theta^{\text{ub}}(2\sum_{i=1}^S K_s)$ are the lower and upper bounds for the combined shell with all points. 
Similarly with~\EEqref{eq:milp_single_c1}, constraints in~\EEqref{eq:milp_multi_c1} and~\EEqref{eq:milp_multi_c2} are respectively for the first and second terms in~\EEqref{eq:milp_multi_cost}. 
$\sum_{s=1}^S h_{s,i} \leq 1$ makes $\uu_i$ to be selected at most in one shell such that the estimated samples are staggered across shells. 
Note that the above MILP in~\EEqref{eq:milp_multi} is for P-D-MS. 
For P-D-MM, \EEqref{eq:milp_multi_c1} and~\EEqref{eq:milp_multi_c2} need to be replaced by~\EEqref{eq:milp_multi_c1_mm} and~\EEqref{eq:milp_multi_c2_mm}
\begin{subequations}\label{eq:milp_multi_mm}
\begingroup\makeatletter\def\f@size{9}\check@mathfonts\def\maketag@@@#1{\hbox{\m@th\normalfont#1}}%
  \begin{align}
  & \arccos(|\uu_{s,i}^T\uu_{s,j}|) \geq \theta_s - (2-h_{s,i}-h_{s,j})M,\ ~\forall s, i>j  \label{eq:milp_multi_c1_mm}\\
  & \arccos(|\uu_{s,i}^T\uu_{s',j}|) \geq \theta_0 - (2-h_{s,i}-h_{s',j})M,\ ~\forall s, s', i>j \label{eq:milp_multi_c2_mm}
  \end{align}
\endgroup
\end{subequations}
where $h_{s,i}=1$ means $\uu_{s,i}$ is selected as one of the $K_s$ points on the $s$-th shell, 
and $\sum_{s=1}^S h_{s,i} \leq 1$ can be removed from~\EEqref{eq:milp_multi_c4}.

Note that although in general the MILP is NP hard, it can be solved with a \emph{globally optimal solution} using branch and bound method which iteratively solves the relaxed LP program.
In our implementation, we solve~\EEqref{eq:milp_single} and~\EEqref{eq:milp_multi} using GUROBI~\cite{gurobi}, 
which can yield the globally optimal solution or at least a reasonable solution within 10 minutes for P-D.  
In practice, we progressively increase the lower bound $\theta^{\text{lb}}$ 
based on the solutions estimated in previous iterations to find a better feasible solution within 10 minutes which is good enough in experiments. 


\section{Experiments}
\label{sec:exp}

\subsection{Effect of Discretization in Incremental Greedy Methods}
\label{sec:exp:discretization}

\begin{figure*}[t!]
  \pgfplotsset{
    myplot style/.style=
    {
      font=\scriptsize,width=0.5\textwidth,
      enlarge x limits=0.15, 
    area legend}}
  \begin{center}
\begin{tabular}{c c}

\begin{tikzpicture} 
  \begin{axis}
    [
      width=0.45\textwidth, 
      xlabel={Number of samples $K$}, ylabel={Covering Radius (degree)}, 
     legend style={at={(0.8,1.0)}, font=\tiny, anchor=north},
      xtick={0,10,...,81},
      ytick={0,10,...,70},
      mark options={scale=1.5,solid},
      grid=major, grid style={dashed},
      xmin=0, xmax=81,
      ymin=10, ymax=70,
      mark size=1.2,
      ytick align=outside,
      xtick align=outside,
      axis y line*=left, 
      axis x line*=bottom
    ] 
    
    \addplot[color=cyan,mark=square*] 
    coordinates { (5, 63.4349) (10,46.6746) (15,38.0275) (20,32.7072) (25,29.1236) (30,26.8192) (35,24.7913) (40,23.2393) (45,22.0477) (50,20.8775) (55,19.8000) (60,19.1644) (65,18.2769) (70,17.6972) (75,16.9653) (80,16.4785) }; 
    \addlegendentry{IMOC+1-Opt+CNLO ($20481$)}
    
    \addplot[color=magenta,mark=square*] 
    coordinates { (5, 63.0626) (10,46.1236) (15,37.3947) (20,31.8891) (25,28.1332) (30,26.1518) (35,24.1227) (40,22.4264) (45,21.2512) (50,20.1648) (55,19.0759) (60,18.2234) (65,17.7028) (70,16.9650) (75,16.4396) (80,15.8378) }; 
    \addlegendentry{IMOC+1-Opt ($20481$)}
    
    \addplot[color=red,mark=square] 
    coordinates { (5, 62.838) (10,45.4968) (15,37.3947) (20,31.8891) (25,28.0063) (30,26.1518) (35,24.0962) (40,22.4046) (45,21.2498) (50,20.1648) (55,19.0759) (60,18.2131) (65,17.6643) (70,16.965) (75,16.4396) (80,15.8376) }; 
    \addlegendentry{IMOC ($20481$)}

    \addplot[color=red,mark=o,dashed] 
    coordinates { (5, 58.2825) (10,42.4238) (15,31.7175) (20,30.4803) (25,18.6994) (30,18.6994) (35,18.6994) (40,18) (45,18) (50,18) (55,15.8587) (60,15.8587) (65,15.8587) (70,15.8587) (75,15.8587) (80,15.8587) }; 
    \addlegendentry{IMOC ($81$)}
    
    \addplot[color=blue,mark=square] 
    coordinates { (5, 54.7223) (10,36.1047) (15,28.8213) (20,27.5785) (25,24.7307) (30,19.3378) (35,18.678) (40,18.2198) (45,18.0862) (50,17.6799) (55,16.141) (60,15.0258) (65,14.6115) (70,13.3309) (75,13.0952) (80,12.8433) }; 
    \addlegendentry{ISC ($20481$)}

    \addplot[color=blue,mark=o,dashed] 
    coordinates { (5,47.5762) (10,35.0876) (15,29.4875) (20,18.6994) (25,18) (30,18) (35,18) (40,16.4125) (45,16.4125) (50,15.8587) (55,15.8587) (60,15.8587) (65,15.8587) (70,15.8587) (75,15.8587) (80,15.8587) }; 
    \addlegendentry{ISC ($81$)}
    
    \addplot[color=green,mark=square] 
    coordinates { (5,48.3704) (10,37.6061) (15,29.3187) (20,25.107) (25,21.7733) (30,20.0606) (35,18.1316) (40,17.0173) (45,16.2243) (50,15.5244) (55,15.0461) (60,14.1783) (65,13.2877) (70,12.4708) (75,12.329) (80,11.8589) }; 
    \addlegendentry{IEEM ($20481$)}
    
    \addplot[color=green,mark=o,dashed] 
    coordinates { (5,46.5129) (10,34.6521) (15,26.565) (20,18.486) (25,16.4124) (30,16.4124) (35,16.4124) (40,15.8587) (45,15.8587) (50,15.8587) (55,15.8587) (60,15.8587) (65,15.8587) (70,15.8587) (75,15.8587) (80,15.8587) }; 
    \addlegendentry{IEEM ($81$)}

    \addplot[color=black,mark=triangle] 
    coordinates { (5,63.4349) (10,46.6746) (15,38.1349) (20,32.7071) (25,29.2486) (30,26.9983) (35,24.8702) (40,23.3293) (45,22.0481) (50,20.8922) (55,20.1034) (60,19.1839) (65,18.3924) (70,17.7739) (75,17.1919) (80,16.6464) }; 
    \addlegendentry{GRASSC}
    
    \addplot[color=black,mark=square] 
    coordinates { (5,59.106) (10,45.9721) (15,36.9502) (20,30.5648) (25,27.861) (30,25.6382) (35,23.1134) (40,22.3934) (45,20.4438) (50,19.4729) (55,18.2012) (60,18.2769) (65,17.5396) (70,16.385) (75,16.1735) (80,15.3308)  }; 
    \addlegendentry{EEM (CAMINO)}
    
    
  \end{axis} 
\end{tikzpicture}

&
\begin{tikzpicture} 
  \begin{axis}
    [
      width=0.45\textwidth, 
      xlabel={Number of samples ($K$ per shell)}, ylabel={Covering Radius (degree)}, 
     legend style={at={(0.9,1.1)}, font=\tiny, anchor=north},
      xtick={0,5,...,27},
      ytick={0,10,...,70},
      mark options={scale=1.5,solid},
      grid=major, grid style={dashed},
      xmin=4, xmax=27,
      ymin=10, ymax=70,
      mark size=1.2,
      ytick align=outside,
      xtick align=outside,
      axis y line*=left, 
      axis x line*=bottom
    ] 
    \addplot[color=cyan,mark=square*] 
    coordinates { (5,61.4046) (10,44.5810) (15,36.7449) (20,29.2180) (25,27.4191)  }; 
    \addlegendentry{IMOC+1-Opt+CNLO ($\theta_s$, $20481$)}
    
    \addplot[color=cyan,mark=triangle*] 
    coordinates { (5,32.9251) (10,23.7532) (15,20.0513) (20,16.6504) (25,15.5103)  }; 
    \addlegendentry{IMOC+1-Opt+CNLO ($\theta_0$, $20481$)}
    
    \addplot[color=magenta,mark=square*] 
    coordinates { (5,58.2496) (10,39.9899) (15,32.8683) (20,27.8956) (25,25.7212)  }; 
    \addlegendentry{IMOC+1-Opt ($\theta_s$, $20481$)}
    
    \addplot[color=magenta,mark=triangle*] 
    coordinates { (5,33.0194) (10,23.2000) (15,19.1083) (20,16.0318) (25,14.8512)  }; 
    \addlegendentry{IMOC+1-Opt ($\theta_0$, $20481$)}
    
    \addplot[color=red,mark=square] 
    coordinates { (5,55.057) (10,38.818) (15,32.194) (20,27.809) (25,25.659)  }; 
    \addlegendentry{IMOC ($\theta_s$, $20481$)}
    
    \addplot[color=red,mark=triangle] 
    coordinates { (5,31.53) (10,22.3258) (15,18.5142) (20,16.0173) (25,14.7783) }; 
    \addlegendentry{IMOC ($\theta_0$, $20481$)}

    \addplot[color=red,mark=o,dashed] 
    coordinates { (5,55.428) (10,31.718) (15,30.480) (20,26.565) (25,18.162)  }; 
    \addlegendentry{IMOC ($\theta_s$, $81$)}
    
    \addplot[color=red,mark=triangle,dashed] 
    coordinates { (5,31.7175) (10,18) (15,18) (20,15.8587) (25,15.8587) }; 
    \addlegendentry{IMOC ($\theta_0$, $81$)}
    
    \addplot[color=blue,mark=square] 
    coordinates { (5,46.113) (10,40.828) (15,25.265) (20,24.156) (25,20.623)  }; 
    \addlegendentry{ISC ($\theta_s$, $20481$)}
    
    \addplot[color=blue,mark=triangle] 
    coordinates { (5,22.5) (10,16.5461) (15,10.9061) (20,10.9029) (25,10.9029) }; 
    \addlegendentry{ISC ($\theta_0$, $20481$)}

    \addplot[color=blue,mark=o,dashed] 
    coordinates { (5,44.533) (10,32.696) (15,20.855) (20,18) (25,16.757)  }; 
    \addlegendentry{ISC ($\theta_s$, $81$)}
    
    \addplot[color=blue,mark=triangle,dashed] 
    coordinates { (5,18) (10,15.8587) (15,15.8587) (20,15.8587) (25,15.8587) }; 
    \addlegendentry{ISC ($\theta_0$, $81$)}
    
  \end{axis} 
\end{tikzpicture}

\end{tabular}
  \end{center}
  \caption{\small\label{fig:exp:discretization}
  \textbf{Effect of discretization}. 
  The left figure shows the covering radii of single shell schemes with $K$ samples obtained by methods using two uniform sets with $81$ samples and $20481$ samples. 
  The right figure shows the covering radii of multi-shell schemes with $K\times 3$ samples, where $\theta_s$ is the mean of covering radii in three shells, and $\theta_0$ is the covering radius of the combined shell. 
} 
\end{figure*}
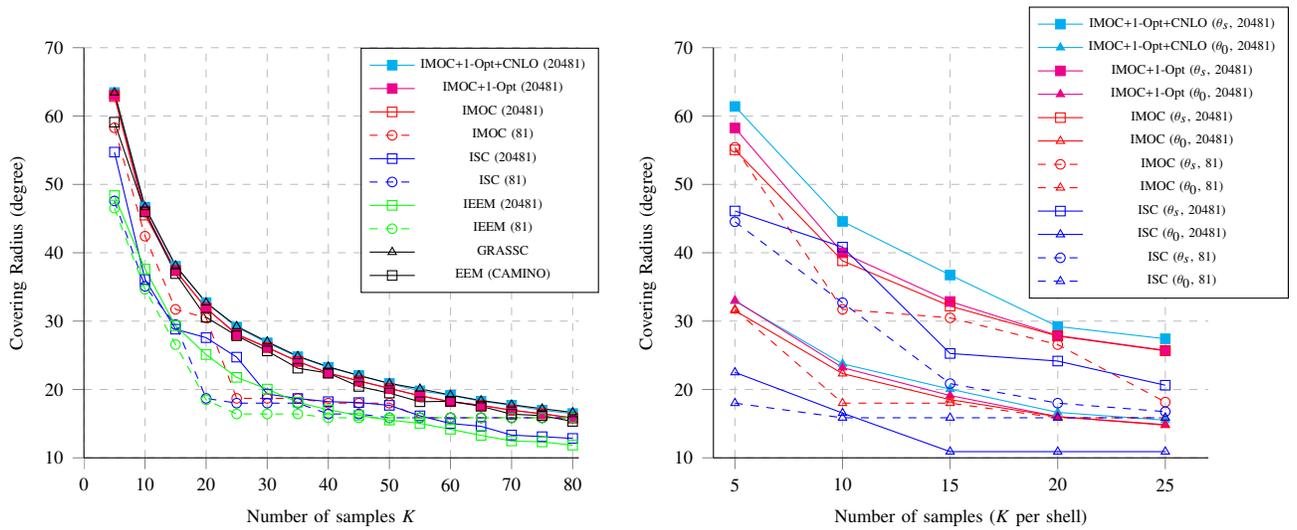

Incremental greedy methods (i.e., IEEM~\cite{deriche_MIA09}, IGEEM~\cite{caruyer_MRM13}, ISC, and IMOC), 
all require a uniform sample set as a discretization of $\mathbb{S}^2$ to determine the best sample in each step. 
It is obvious that a finer uniform set may not yield better results for ISC in 2D case. 
We test the discretization effect in $\mathbb{S}^2$ for both single shell and multi-shell scheme design. 
We consider two uniform sample sets respectively with $81$ samples and $20482$ samples from spherical tessellation. 
With these two uniform sets, IEEM, ISC, IMOC, and IMOC + 1-Opt are performed to generate single shell scheme with $K\in[5,80]$, 
where IMOC + 1-Opt means 1-Opt with the initialization by IMOC.
The left subfigure of Fig.~\ref{fig:exp:discretization} depicts the covering radii of schemes obtained by different methods with different uniform sets. 
It also shows the best known single shell schemes using the SC concept collected by Dr.\ Sloane~\textsuperscript{\ref{fn:neil}}, denoted as GRASSC, and the best known single shell schemes using EEM collected in CAMINO~\cite{cook_ISMRM06}. 
When $K\in[50,80]$ is close to $81$ (the number of uniform samples in the coarse uniform set), covering radii by IEEM and ISC are not improved using the finer set, 
and when $K<50$ is far from $81$, IEEM and ISC obtain larger covering radii using finer discretization. 
When using the finer uniform set, covering radii by IMOC and IMOC + 1-Opt are improved, close to the best known solutions in GRASSC, 
and are much larger than covering radii of the schemes in CAMINO. 
ISC, IMOC, IMOC + 1-Opt are performed to generate schemes with 3 shells, $K\in[5,25]$ per shell by using these two uniform sets. 
The right subfigure of Fig.~\ref{fig:exp:discretization} shows mean of covering radii of three shells and the covering radii of the combined shell. 
It also demonstrates that when $K\times 3$ is close to $81$, covering radii of the combined shell by ISC are not improved using the finer set, 
while IMOC has no such issue. 
The experiment shows that IMOC and IMOC + 1-Opt generally obtain better results than ISC and IEEM for P-C. 
CNLO can further improve the results by IMOC + 1-Opt, 
and the covering radii by IMOC + 1-Opt + CNLO obtains the best results, very closed to the solutions in GRASSC in single shell case. 
Thus, as shown in Table~\ref{tab:algorithm_problems_2}, ISC is suggested for P-O, and IMOC + 1-Opt + CNLO is suggested for P-C.

\subsection{Covering Radii in Sampling Schemes}
\label{sec:exp:CR}

\begin{figure*}[t!]
  \begin{center}
  \scalebox{0.93}{
    \begin{tabular}{ c c c}
\includegraphics[width=0.32\textwidth]{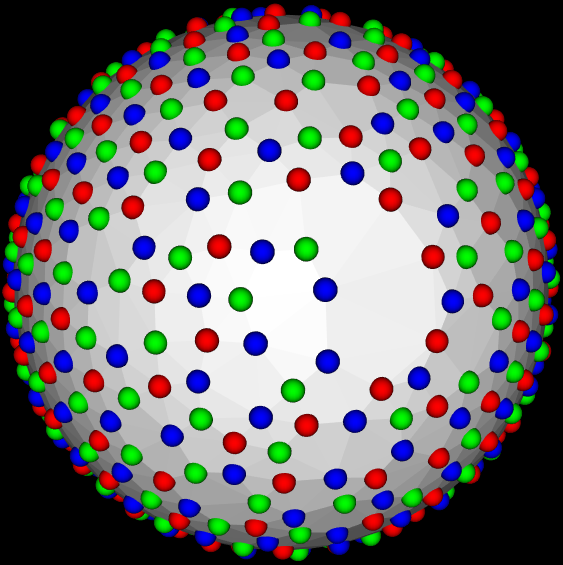} &
      \includegraphics[width=0.32\textwidth]{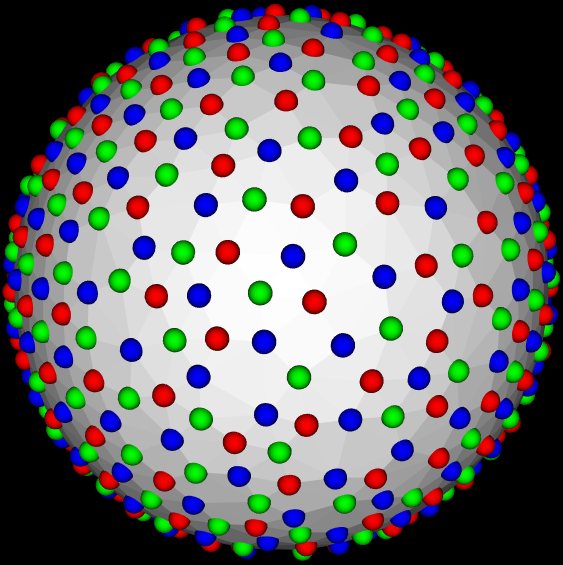}  &
\includegraphics[width=0.32\textwidth]{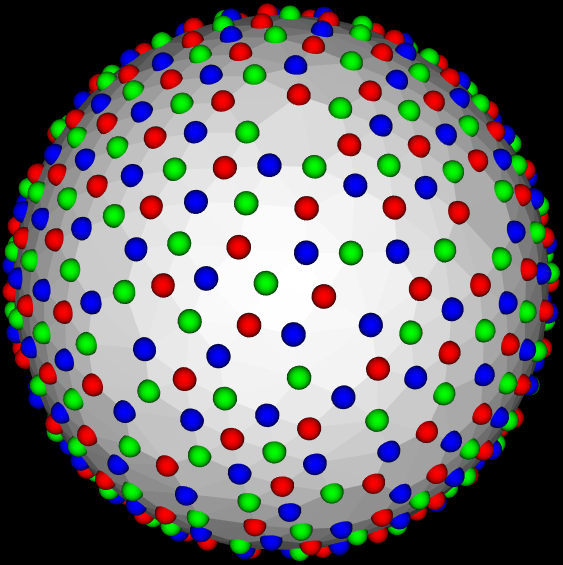} \\
IMOC  & IMOC + 1-Opt   & IMOC + 1-Opt + CNLO  \\
      ($13.49^\circ$, $13.48^\circ$, $13.49^\circ$, $7.78^\circ$)  & ($13.56^\circ$, $13.50^\circ$, $13.50^\circ$, $7.78^\circ$)  & ($\textbf{14.56}^\circ$, $\textbf{14.64}^\circ$, $\textbf{14.69}^\circ$, $\textbf{8.40}^\circ$) 
    \end{tabular}
  }
  \end{center}
  \vspace{-3mm}
    \caption{\label{fig:visual}Multi-shell sampling schemes with $90\times 3$ samples generated by three methods, and the covering radii of 3 shells and the combined shell in schemes shown in Table~\ref{tab:grad_28}. 
    The colors differentiate the sampling points from the three shells. Note that there is a hole area in the scheme by IMOC, which can be fixed by 1-Opt and CNLO.}
  \vspace{-4mm}
\end{figure*}

We evaluate the proposed method by generating multi-shell schemes with $K\times 3$ samples. 
We test two cases with $K=28$ and $K=90$, 
considering these two cases were also used for evaluation in~\cite{caruyer_MRM13}, 
and the multi-shell scheme with $90\times 3$ samples by IGEEM was also used in HCP~\cite{caruyer_MRM13,sotiropoulos_HCP_NI13}. 

Table~\ref{tab:grad_28} shows the covering radii of the schemes by different methods, 
where IMOC + 1-Opt + CNLO means CNLO using the result by IMOC + 1-Opt as the initialization. 
Sparse and Optimal Acquisition (SOA) in~\cite{koay_MP2012}, which separates a uniform scheme into several sets by using a greedy algorithm, 
is performed using the software released by the author~\footnote{\url{https://sites.google.com/site/hispeedpackets/Home/soa-design}}. 
The result by SOA for the scheme with $90\times 3$ samples is not shown, because it did not terminate after running for 2 days. 
The solid rotation method in~\cite{desatis:ismrm2011} starts from the single shell scheme with $28$ samples by EEM, and finds the optimal three rotations for the 3 shells, such that the electrostatic energy in the combined shell is minimized. 
The scheme by the solid rotation method is the 3 rotated EEM schemes, and the covering radii of the scheme were extracted from~\cite{caruyer_MRM13}. 
Note that we run GEEM by using the latest software released by the authors~\footnote{\url{https://github.com/ecaruyer/qspace}}, 
which obtains better covering radii for individual shells but a slightly worse covering radius for the combined shell than the original results shown in~\cite{caruyer_MRM13}. 
%
As references, the table also lists the best known results for single shell with $K$ and $K\times 3$ samples using EEM from CAMINO~\cite{cook_ISMRM06} and the SC formulation from~\textsuperscript{\ref{fn:neil}}~\cite{conway_packing_1996}. 
Note that the precomputed schemes by EEM and GRASSC are only for the single shell case, not for the multi-shell case. 
If this single shell scheme (with $28$ or $90$ samples) is repeated for 3 shells (i.e., with 3 different b values), then the covering radius for the combined shell (with $84$ samples or $270$ samples) is actually $0^\circ$. 
The scheme with $84$ samples by EEM is not related with the scheme with 28 samples by EEM. 
IMOC obtains better covering radii than incremental methods, i.e., IGEEM, ISC, and it is even better than MILP, 
which is because the globally optimal schemes by MILP are determined only from a coarse uniform set with $321$ samples, not from a fine uniform set. 
It is impractical for MILP to work with $20481$ samples, because MILP is NP hard. 
For both $28\times 3$ and $90\times 3$ cases, 
IMOC + 1-Opt + CNLO obtains best schemes with largest covering radii in both 3 individual shells and the combined shell, 
and the covering radii by IMOC + 1-Opt + CNLO in these 3 shells are even better than best known scheme collected in CAMINO using EEM~\cite{jones_MRM99,cook_ISMRM06}, 
similarly with the left subfigure in Fig.~\ref{fig:exp:discretization}. 

Table~\ref{tab:grad_28} also shows the running time for different methods on an ordinary laptop. 
Incremental methods (IGEEM, ISC, IMOC) are all very fast. 
IGEEM and ISC have similar running time, because they can be implemented in the same efficient way to maintain the table of increased cost functions between the selected sample set and each unselected sample, 
and update the table when each new sample is selected. 
A similar acceleration strategy can be used for 1-Opt. 
MILP is terminated after 10 minutes, because we found that the cost function of MILP changes very slow after 10 minutes. 
Compared with the $28\times 3 $ scheme, the running time of estimating the $90\times 3$ scheme by all methods except IMOC increases largely. 
The running time of GEEM increases from $3.7$ seconds to $11$ minites, and the running time of CNLO increases from $39.7$ seconds to $31$ minutes, because of the larger number of constraints. 
The running time of IMOC decreases for the $90\times 3$ scheme, probably because the iteration number of MOC decreases for the $90\times 3$ scheme, compared with the $28 \times 3$ scheme. 
It should be noted that the running time of scheme design methods is actually not very important for practical usage, because the schemes are optimized only once for scanners.

Fig.~\ref{fig:visual} shows the schemes by IMOC, IMOC + 1-Opt, and IMOC + 1-Opt + CNLO. 
It shows that 
1) IMOC + 1-Opt can fix the hole in the scheme by IMOC; 
2) IMOC + 1-Opt increases the covering radii of individual shells of the scheme by IMOC, although the covering radius of the combined shell keeps the same ($7.78^\circ$);
3) CNLO can refine the scheme by IMOC + 1-Opt. 
Thus for P-C, we do not use IMOC as a stand-alone method. 
Instead, IMOC + 1-Opt + CNLO is suggested for P-C, where two greedy methods (i.e., IMOC + 1-Opt) are used to set initialization of gradient based optimization method CNLO. 

\begin{table*}[t!]
  \caption{\label{tab:grad_28}\small Covering radii of multi-shell sampling schemes with $28\times 3$ and $90\times 3$ samples generated by various methods with the corresponding running time. 
}
  \centering
  \begin{tabular}{ c | c | c | c | c | c}
    \hline 
    & Shell 1 (28)  & Shell 2 (28)  & Shell 3 (28) & Combined ($28\times 3$)  & Running Time\\
    \hline   
    SOA~\cite{koay_MP2012} & $15.9^\circ$ & $15.8^\circ$ & $16.7^\circ$  & $15.1^\circ$  & 36 min\\
    Solid Rotation~\cite{desatis:ismrm2011} & $25.7^\circ$ & $25.7^\circ$ & $25.7^\circ$  & $7.4^\circ$  & not available\\
    IGEEM~\cite{caruyer_MRM13} & $19.2^\circ$ & $19.7^\circ$ & $19.3^\circ$ & $4.7^\circ$ & 0.3 s\\
    GEEM~\cite{caruyer_MRM13} & $23.6^\circ$ & $24.3^\circ$ & $23.4^\circ$ & $12.3^\circ$ & 3.7 s\\
    ISC ($N=20481$)  & $21.3^\circ$ & $19.3^\circ$ & $21.1^\circ$ & $10.5^\circ$  & 0.3 s\\
    MILP ($N=321$)  & $23.8^\circ$ & $23.8^\circ$  & $24.3^\circ$ & $13.3^\circ$  & 10 min \\
    IMOC ($N=20481$) & $24.3^\circ$ & $24.3^\circ$ & $24.3^\circ$ & $14.0^\circ$ & 10.5 s \\
    IMOC + 1-Opt ($N=20481$) & $24.3^\circ$ & $24.4^\circ$ & $24.3^\circ$ & $14.0^\circ$ &  12.8 s \\
    IMOC + 1-Opt + CNLO & $\textbf{26.1}^\circ$ & $\textbf{26.3}^\circ$ & $\textbf{26.9}^\circ$ & $\textbf{14.4}^\circ$ & 39.7 s \\
    \hline
    EEM (CAMINO)~\cite{jones_MRM99,cook_ISMRM06} & $25.7^\circ$ & $25.7^\circ$ & $25.7^\circ$ & $15.6^\circ$ & not available\\
    GRASSC~\textsuperscript{\ref{fn:neil}}~\cite{conway_packing_1996} & $27.8^\circ$ & $27.8^\circ$ & $27.8^\circ$ & $16.2^\circ$ & not available\\
    \hline 
  \hline 
       & Shell 1 (90)  & Shell 2 (90)  & Shell 3 (90) & Combined ($90\times 3$) & Running Time\\
  \hline   
    IGEEM \cite{caruyer_MRM13} & $10.8^\circ$ & $10.3^\circ$ & $10.5^\circ$ & $2.4^\circ$ & 0.6 s \\
    GEEM~\cite{caruyer_MRM13} & $12.9^\circ$ & $13.3^\circ$ & $12.6^\circ$ & $7.2^\circ$ & 11 min \\
    ISC ($N=20481$)  & $10.4^\circ$ & $9.7^\circ$ & $10.4^\circ$ & $4.6^\circ$  & 0.6 s \\
    MILP ($N=321$)  & $13.3^\circ$ & $13.5^\circ$  & $13.3^\circ$ & $7.9^\circ$  & 10 min \\
    IMOC ($N=20481$) & $13.5^\circ$ & $13.5^\circ$ & $13.5^\circ$ & $7.8^\circ$ & 7.8 s \\
    IMOC + 1-Opt ($N=20481$) & $13.6^\circ$ & $13.5^\circ$ & $13.5^\circ$ & $7.8^\circ$ & 29.6 s\\
    IMOC + 1-Opt + CNLO & $\textbf{14.6}^\circ$ & $\textbf{14.6}^\circ$ & $\textbf{14.7}^\circ$ & $\textbf{8.4}^\circ$ & 31 min\\
  \hline
    EEM (CAMINO)~\cite{jones_MRM99,cook_ISMRM06} & $15.1^\circ$ & $15.1^\circ$ & $15.1^\circ$ & not available & not available \\
    GRASSC~\textsuperscript{\ref{fn:neil}}~\cite{conway_packing_1996} & $15.7^\circ$ & $15.7^\circ$ & $15.7^\circ$ & not available & not available \\
  \hline
\end{tabular}

\end{table*}

\subsection{Rotational Invariance in Reconstruction}
\label{sec:exp:reconstruction}

We test rotational invariance of the schemes with $28\times 3$ samples by different methods in Table~\ref{tab:grad_28}. 
We generate synthetic diffusion signal from a mixture of cylinder model~\cite{OzarslanNeuroImage2006}, 
where two cylinders cross with $60^\circ$ and share the cylinder parameters of length $L=\SI{5}{mm}$, radius $\rho=\SI{5}{\micro\metre}$, diffusivity $D_0= \SI{2.02e-3}{mm^2/s}$. 
With each tested scheme, we rotate the model and generate signals $20481$ times with the rotation angles determined by the uniform sample set with $20481$ samples. 
Then we perform Non-Negative Spherical Deconvolution (NNSD)~\cite{cheng_NI2014} with the spherical harmonic (SH) order $8$ to estimate fODFs, detect the peaks, 
and calculate the mean angular differences by comparing the detected peaks with the ground-truth fiber directions in these $20481$ tests. 
A good sampling scheme should give low mean angular differences with low standard deviations. 
Table~\ref{tab:estimation} lists the mean and standard deviation of angular differences obtained by NNSD using different schemes. 
IMOC + 1-Opt +CNLO yields the lowest angular differences with the lowest deviation, and IMOC has the lower mean and deviation than IGEEM. 

We compare the $28\times 3$ schemes and $90\times 3$ schemes in DWI signal reconstruction by using tensorial Spherical Polar Fourier Imaging (SPFI)~\cite{cheng_MICCAI2015}. 
The DWI signals are synthesized using a mixture of tensor model with two symmetric tensors with eigenvalues $[1.7,0.2,0.2]\times 10^{-3}\,\text{mm}^2/\text{s}$ with a crossing angle of $60^\circ$. 
We rotate the model and generate signals $20481$ times with the rotation angles determined by the uniform sample set with $20481$ samples. 
Tensorial SPFI with spherical order of $8$ and radial order of $4$ is used to fit the data and re-sample the DWI signals at continues 3 shells with a fine tessellation of $20481$ samples. 
Then normalized mean Root-Mean-Squared-Error (RMSE) of DWI signals in 3 shells is calculated. 
We also add Rician noise with SNR=$30$ (i.e., standard deviation of $1/30$) for 1000 trails, and re-calculate mean normalized RMSE in reconstruction using noisy samples. 
Table~\ref{tab:estimation_spf} shows the mean and standard deviation of normalized RMSE in reconstruction using noise-free samples and noisy samples. 
It shows that IMOC + 1-Opt + CNLO obtains lower mean and std of normalized RMSE values in reconstruction, especially when the number of samples is small and SNR is high. 
However, when using a larger number of noisy samples in SPFI (e.g., $90\times 3$ scheme with noise, SNR=30), the normalized RMSE results by IMOC + 1-Opt + CNLO and GEEM are similar. 
It also shows that the $90\times 3$ scheme by IMOC has much larger standard deviation (i.e., rotational invariance) of normalized RMSE than schemes by other methods, which is because of the hole area effect in IMOC. 
IMOC + 1-Opt and IMOC + 1-Opt + CNLO reduce the rotational invariance of IMOC.

\begin{table*}[t!]\footnotesize
  \caption{\label{tab:estimation}Angular differences between estimated directions and ground-truth fiber directions using different schemes by various methods.}
  \centering
    \begin{tabular}{ c|  c | c | c | c | c | c}
      \hline
    & SOA~\cite{koay_MP2012}  &  IGEEM~\cite{caruyer_CDMRI11,caruyer_MRM13} & GEEM~\cite{caruyer_MRM13}  & IMOC & IMOC + 1-Opt & IMOC + 1-Opt + CNLO \\
      \hline
      Angular Difference & $1.41^\circ \pm 0.75^\circ$  & $1.67^\circ \pm 0.81^\circ$  & $1.23^\circ \pm 0.73^\circ$  & $1.43^\circ \pm 0.77^\circ$ & $1.31^\circ \pm 0.71^\circ$ & $\textbf{1.16}^\circ \pm \textbf{0.64}^\circ$   \\
      \hline
    \end{tabular}
\end{table*}

\begin{table*}[t!]\footnotesize
  \caption{\label{tab:estimation_spf}Normalized RMSE of DWI signal reconstruction using tensorial SPFI.}
  \centering
    \begin{tabular}{ c|  c | c | c | c | c | c}
      \hline
                         & SOA~\cite{koay_MP2012}  &  IGEEM~\cite{caruyer_CDMRI11,caruyer_MRM13} & GEEM~\cite{caruyer_MRM13}  & IMOC & IMOC + 1-Opt & IMOC + 1-Opt + CNLO \\
      \hline
      $28\times 3$, no noise &  $0.46\% \pm 0.048\%$ & $0.42\% \pm 0.046\%$  & $0.38\% \pm 0.062\%$  & $0.38\% \pm 0.0058\%$ & $0.36\% \pm \textbf{0.054}\%$ & $\textbf{0.35\%} \pm 0.055\%$   \\
      $90\times 3$, no noise &  not available & $0.176\% \pm 0.0044\%$  & $0.174\% \pm 0.0022\%$  & $0.179\% \pm 0.016\%$ & $0.176\% \pm 0.0035$ & $\textbf{0.172\%} \pm \textbf{0.0021}\%$   \\
      $28\times 3$, SNR=$30$ &  $4.39\% \pm 0.58\%$ &  $4.33\% \pm 0.60\%$  & $\textbf{4.31}\% \pm 0.58\% $  & $4.35\% \pm 0.61\% $ & $4.33\% \pm 0.59\%$ & $\textbf{4.31}\% \pm \textbf{0.56}\% $   \\
      $90\times 3$, SNR=$30$ &  not available &  $2.57\% \pm 0.34\%$ & $2.55\% \pm \textbf{0.33}\% $  & $2.58\% \pm 0.35\% $ & $2.56\% \pm 0.34\%$ & $\textbf{2.54}\% \pm \textbf{0.33}\%$  \\
      \hline
    \end{tabular}
\end{table*}

\subsection{MILP for Separation of Sampling Schemes} 

We evaluate the effectiveness of the proposed MILP in~\EEqref{eq:milp_multi} with $w=1$ to solve P-D-MS, 
by verifying whether it can separate a set of samples into several subsets, keeping the points in each subset as uniform as possible. 
We compared MILP with the \texttt{subsetpoints} program in CAMINO~\cite{jones_MRM99,cook_ISMRM06} and \texttt{dirsplit} in MRtrix~\footnote{\url{http://mrtrix.readthedocs.io/en/latest/reference/commands/dirsplit.html}} for the same task. 
%
For this evaluation, we randomly mix two sets of uniform points, 
where one set has $81$ points generated by spherical tessellation (with covering radius $15.9^\circ$) 
and the other set has $60$ points (with covering radius $18.3^\circ$) from CAMINO generated by EEM. 
Separation of these $141$ points into subsets respectively with $81$ and $60$ samples should ideally give results that match the original uniform point sets. 
MILP obtained results that exactly match the original point sets within 1 second. 
\texttt{subsetpoints} in CAMINO, which minimizes electrostatic energy by using simulated annealing, saves the result every hour when it runs. 
It obtains two incorrect points in each subset after running for 2 hours, 7 incorrect points after 8 hours, and the correct result after 9 hours. 
Although the correct result has been obtained, the program continues to run for hours until the simulated annealing temperature is finally small enough.  
%
\texttt{dirsplit} in MRtrix initializes two sets and then randomly swaps 2 directions in each step to reduce the electrostatic energy.  
\texttt{dirsplit} can only separate the scheme into 2 schemes with nearly equal numbers 71 and 70. 
With default parameters and multi-threading, it finished running after 16 minutes. 
However, the covering radii in these two sets are $6.4^\circ$ and $5.1^\circ$ respectively, different from the original two uniform sets. 

\subsection{MILP for sub-sampling of multi-shell HCP scheme} 

\begin{figure*}[t!]
  \begin{center}
  \scalebox{0.95}{
    \begin{tabular}{ c c c}
      HCP scheme, IGEEM ($90\times 3$, $2.4^\circ$) & MILP ($45\times 3$, $\textbf{6.9}^\circ$)  & random sub-sampling ($45\times 3$, $2.4^\circ$)\\
\includegraphics[width=0.28\textwidth]{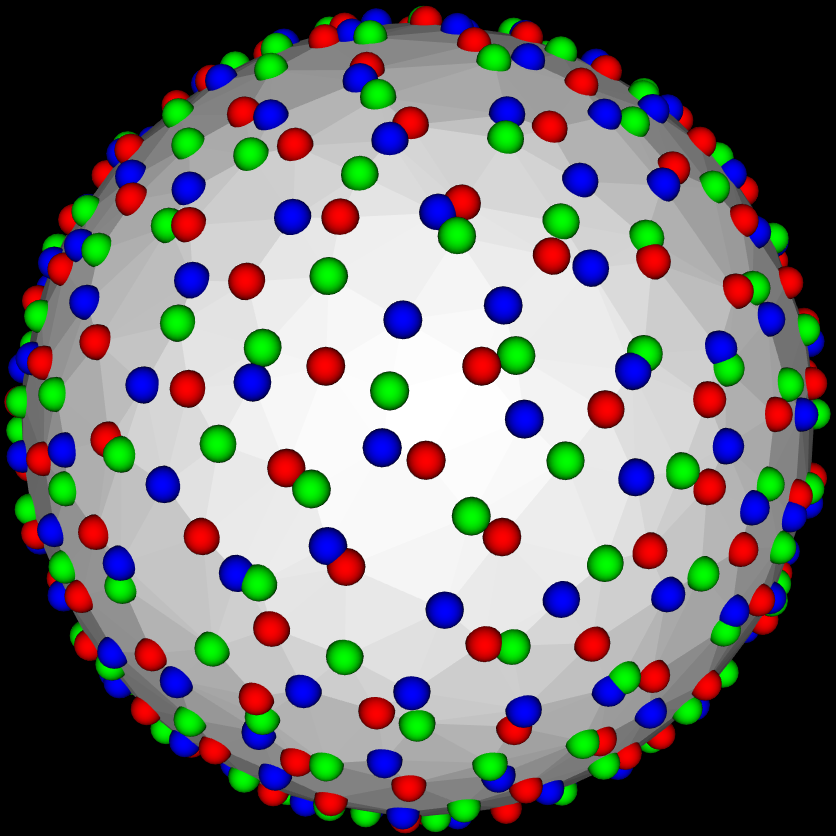} &
      \includegraphics[width=0.28\textwidth]{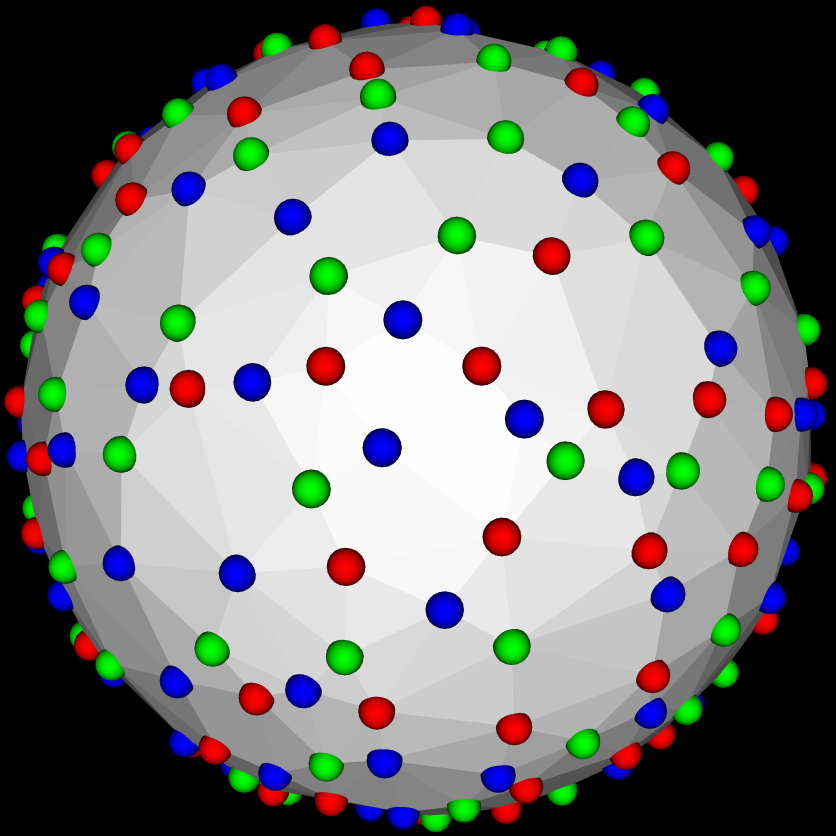}  &
\includegraphics[width=0.28\textwidth]{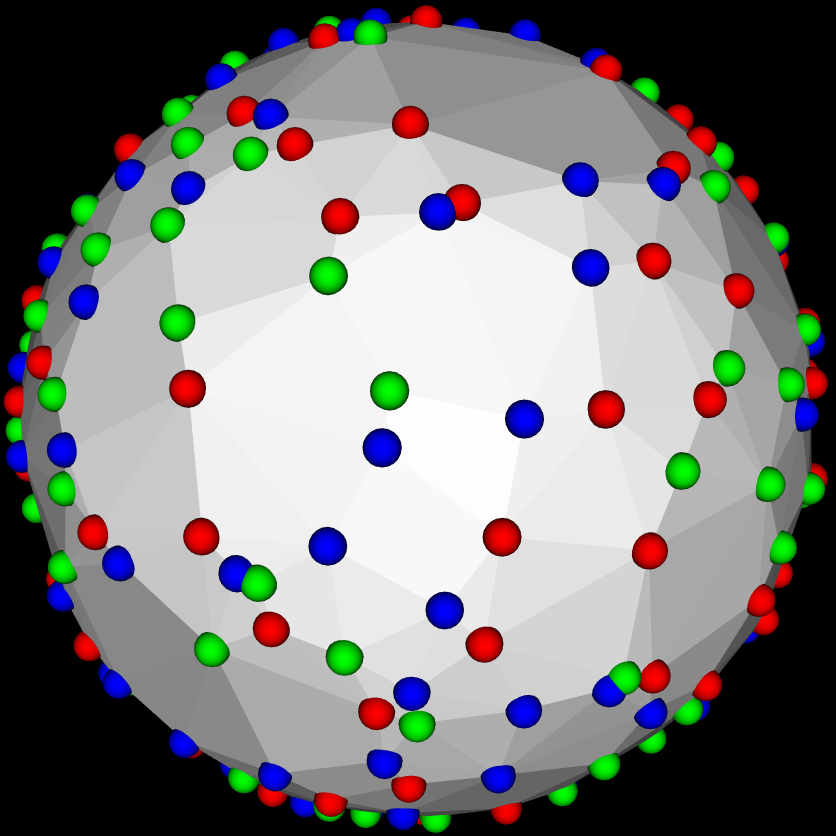} \\
\includegraphics[width=0.28\textwidth]{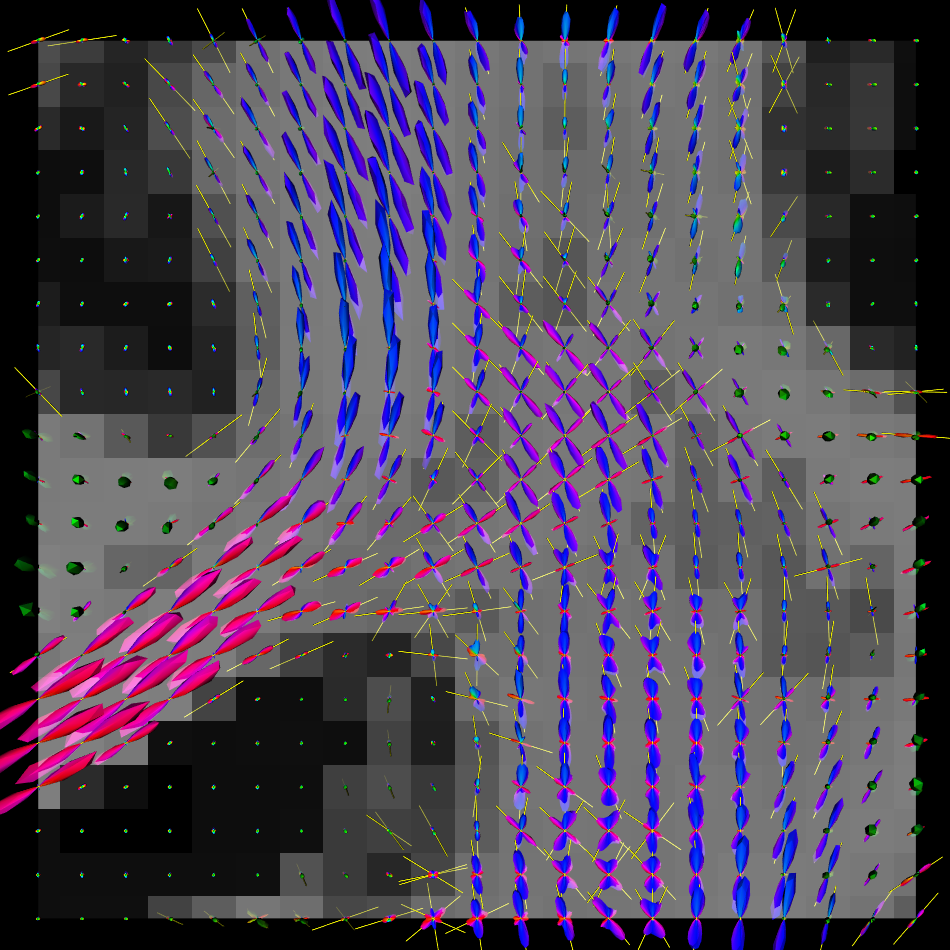} &
      \includegraphics[width=0.28\textwidth]{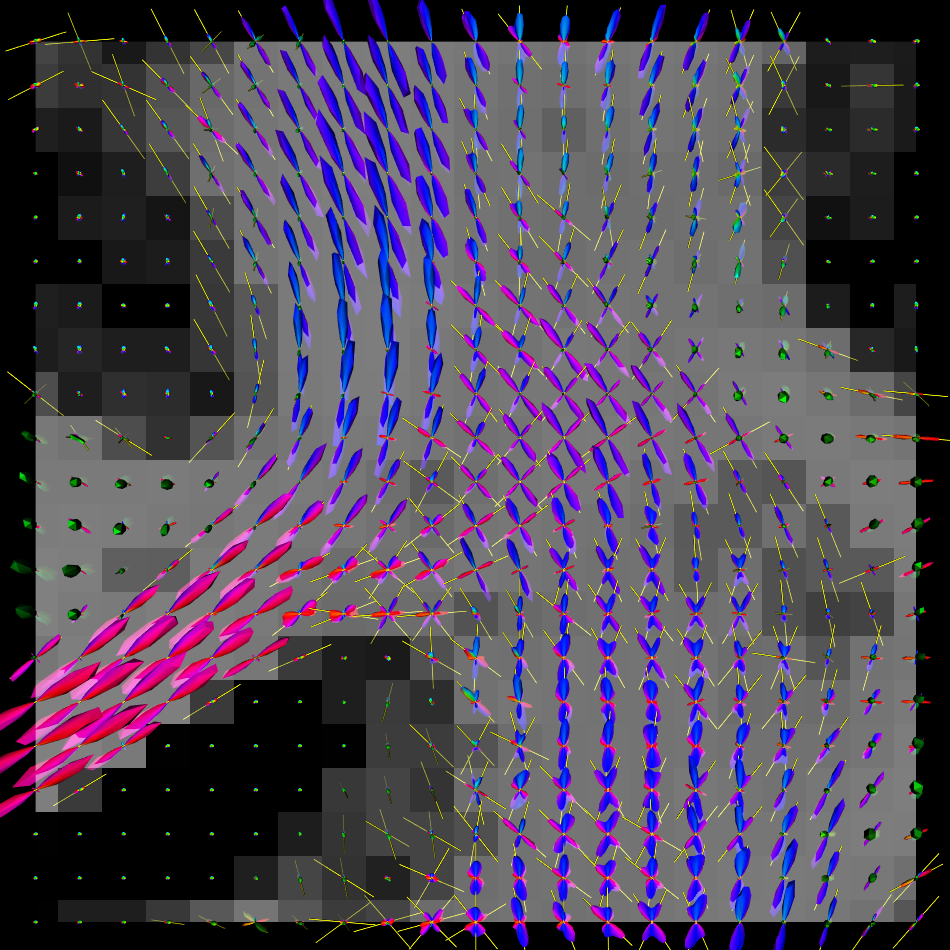}  &
\includegraphics[width=0.28\textwidth]{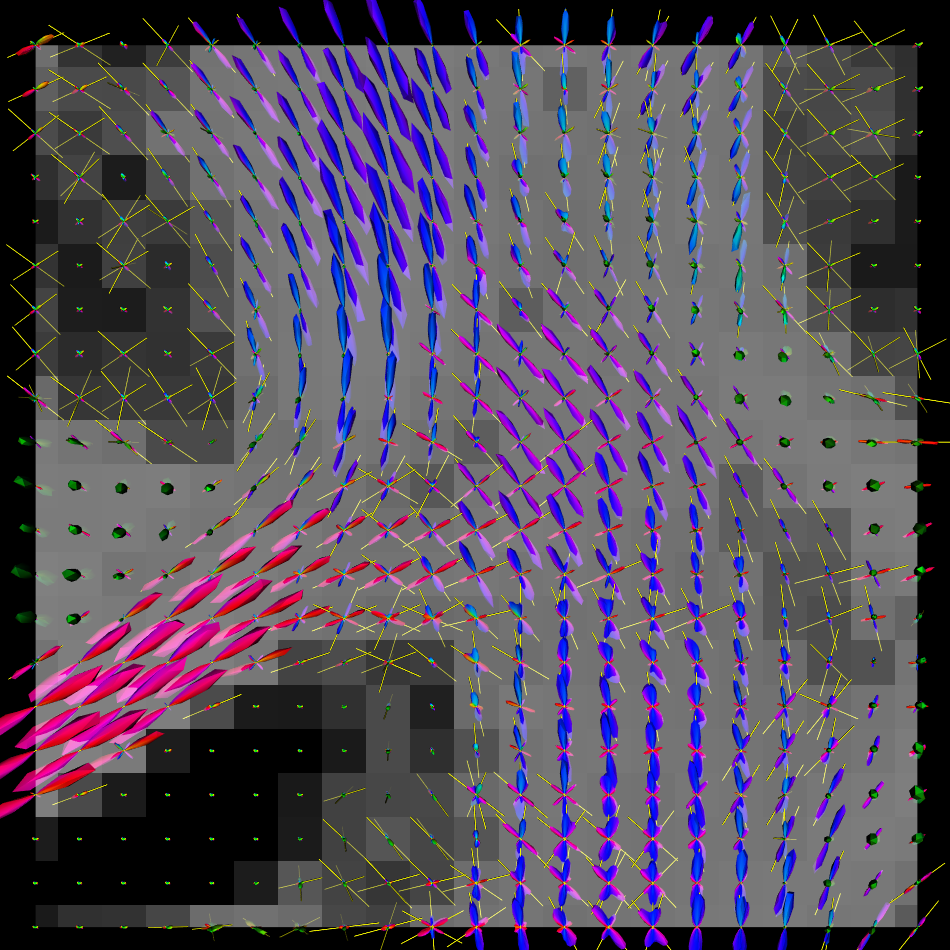} \\
    \end{tabular}
  }
  \end{center}
  \vspace{-3mm}
    \caption{\label{fig:hcp_subsampling}First row: full HCP scheme, sub-sampled schemes by MILP and random selection, respectively.
    Second row: fODFs fields estimated from the data sets corresponding to schemes, 
    where fODF glyphs are colored by directions, the background is the GFA map, and the yellow tubes denote detected peaks. }
  \vspace{-4mm}
\end{figure*}

In order to reduce the scanning time and reuse the existing data, we may want to uniformly select a sub-sampled scheme from an existing scheme. 
IGEEM has been successfully used in multi-shell diffusion data acquisition in HCP~\cite{caruyer_CDMRI11,caruyer_MRM13,sotiropoulos_HCP_NI13}. 
The scheme has 3 shells with b values $1000$, $2000$, and $3000$ $\si{s/mm^2}$, 90 samples per shell. 
See top left subfigure in Fig.~\ref{fig:hcp_subsampling}. 
MILP and a random selection are performed to obtain two sub-sampled schemes with $45\times 3$ samples. 
It is a problem of P-D-MM, which to our knowledge can not be solved by any existing released routine (e.g., in CAMINO or MRtrix). 
MILP yields the covering radius $7.0^\circ$ in the combined shell, while the random selection yields $2.4^\circ$, which is the same as the original HCP scheme. 
See first row in Fig.~\ref{fig:hcp_subsampling}. 
We also perform NNSD with SH order $8$ on a subject with the full scheme of $90\times 3$ samples and two sub-sampled schemes with $45\times 3$ samples from MILP and random selection. 
The corresponding fODF fields with detected peaks are displayed in Fig.~\ref{fig:hcp_subsampling} with GFA as backgrounds.  
Peaks are detected from the fODFs with GFA higher than $0.5$. 
Fig.~\ref{fig:hcp_subsampling} shows that the estimated fODF and peaks from sub-sampled scheme by MILP are much closer to the results from the full scheme, compared with the random selection. 
Note that fODFs near grey matter areas by the scheme from random selection are very noisy, compared with the results using the full scheme. 
That is because that random selection cannnot keep the uniformity in the sub-sampled scheme. 
In the region visualized in Fig.~\ref{fig:hcp_subsampling}, 
the mean angular difference of the detected peaks is $4.2^\circ$ between the results from the full scheme and the sub-sampled scheme by MILP, 
and $5.9^\circ$ between the results from the full scheme and the sub-sampled scheme by random selection. 
The normalized RMSE of the estimated GFA of fODFs is $7.5\%$ for the sub-sampled scheme by MILP and is $18.9\%$ for the scheme by random selection.  

\section{Discussion}

\subsection{Sampling scheme design without antipodal symmetry}
\label{sec:symmetry}
The SC formulation and EEM formulation~\cite{jones_MRM99,caruyer_CDMRI11,caruyer_MRM13,zhan:ISBI2011,koay_MP2012,koay_MRM2014,dubois:2006,cook_JMRI2007} 
both are to design sampling schemes with the antipodal symmetry constraint, 
based on prior knowledge that the diffusion signal is antipodal symmetric, i.e., $E(\q)=E(-\q)$. 
However, some works proposed to model asymmetry of diffusion signals~\cite{Liu_MRM2004} or fiber ODFs~\cite{reisert:TMI2012}, 
which may prefer sampling schemes without the antipodal symmetry constraint. 
Moreover general uniform single or multiple shell sampling schemes (i.e., without the antipodal symmetry constraint) are also be useful in some applications other than diffusion MRI. 

Although the proposed SC formulation and 5 algorithms focus on sampling scheme design with antipodal symmetry, 
they can be easily performed to design general sampling schemes without the antipodal symmetry constraint. 
Actually the original SC formulation in mathematics~\cite{toth_1949,musin_2015} does not consider the antipodal symmetry constraint. 
If the absolute value operator is removed from the definition of covering radius in~\EEqref{eq:CoveringRadius}, 
then~\EEqref{eq:SC_singleShell} and~\EEqref{eq:SC_multiple} are SC formulations respectively for general single and multiple shell scheme design, 
where~\EEqref{eq:SC_singleShell} is the original SC formulation in~\cite{toth_1949,musin_2015}. 
ISC and 1-Opt can be used for general uniform sampling scheme design straightforwardly, based on the revised definition of covering radius. 
For IMOC, if we revise the definition of the coverage set of a point $\Vx$ as $C(\Vx,\theta)= \{\Vy \mid \arccos(\Vy^T\Vx) < \theta, \Vy\in\mathbb{S}^2 \}$, without the absolute value operator, 
then IMOC can be used for general uniform scheme design. 
In Fig.~\ref{fig:IMOC_search} (a), with the revised definition, the covering set does not include the yellow arc generated by the dashed samples in the antipodal positions. 
For CNLO and MILP, if we remove the absolute operator in Eq.~\eqref{eq:CNLO_single},~\eqref{eq:CNLO_single_local},~\eqref{eq:CNLO_multiple},~\eqref{eq:milp_single} and~\eqref{eq:milp_multi}, 
then CNLO and MILP can be used for general uniform sampling scheme design without the antipodal symmetry constraint. 
For each proposed algorithm, the strategy and procedure remain the same if the antipodal symmetry constraint is removed.

\subsection{Sampling scheme design with other considerations}

\subsubsection{Diffusion models and model fitting}
As discussed in the introduction part, 
both the SC formulation and EEM formulation are to design uniform sampling schemes in single or multiple shells with parameters $S$ and $\{K_s\}_{s=1}^S$. 
This is based on the fact that diffusion signals in voxels have various preferred orientations, while we need to design one scheme for all signals in human brain. 
The number of b values $S$, and the number of samples in each shell $\{K_s\}_{s=1}^S$, are both user determined. 
Different diffusion models, e.g., tensor model and various HARDI models, may prefer different distributions of b values in the radial part of the $\q$-space, and different number of samples in low or b values. 
Moreover, even for the same model, different model fitting algorithms may prefer different schemes~\cite{cheng:handbook2016}. 
The samples with high b values capture more orientational information, while have lower single-to-noise ratio (SNR), than the samples with low b values. 
Thus robust estimation algorithms~\cite{chang:MRM2005,cheng_NI2014,cheng_MICCAI2015}, which are less sensitive to noise, may prefer larger number of samples with high b values.

\subsubsection{Eddy current correction}
Recent works on eddy current correction~\cite{irfanoglu:NI2015,andersson:NI2016,graham:NI2016} showed that three kinds of acquisitions are useful for eddy current correction, 
i.e., 1) the blip-up blip-down acquisition, which scans the the data twice with flipped phase-encoding directions;
2) the acquisition with gradients on the whole sphere that includes both $\{\uu_i\}$ and $\{-\uu_i\}$; 
3) the acquisition with gradients on the whole sphere without the antipodal symmetry constraint. 
Schemes by SC can be directly used in the blip-up blip-down acquisition with flipped phase-encoding directions. 
and can be used in the second kind of acquisition by incorporating the flipped gradients $\{-\uu_i\}$. 
For the third acquisition on the whole sphere without the antipodal symmetry constraint, we can perform the proposed optimization methods without the constraint, as described in Section~\ref{sec:symmetry}. 

\subsubsection{Gradient reorientation}
The uniform schemes generated by SC or EEM are used to scan DWI data. 
Some preprocessing steps, including motion correction, eddy current correction, registration of DWIs, etc., may require gradient reorientation~\cite{Alexander_TMI2001,leemans:MRM2009}. 
Gradient reorientation changes each individual gradient, and also the separation angles between gradients. 
Thus even though the designed schemes by SC for scanning data are optimized with large angular separation, the reorientated scheme for model fitting may be sub-optimal. 
This issue also exists for schemes by EEM, where the electrostatic energy is sub-optimal after reorientation. 
The gradient reorientation issue has not been considered in existing works on uniform scheme design~\cite{jones_MRM99,caruyer_CDMRI11,caruyer_MRM13,zhan:ISBI2011,koay_MP2012,koay_MRM2014,dubois:2006,cook_JMRI2007}, 
because the transform matrix for each gradient vector is dependent on DWI images and also the image registration algorithm used in the preprocessing steps. 
In this paper, we show that in both SC and EEM formulations, uniform sampling scheme design before gradient reorientation is an approximation of uniform sampling scheme design after gradient reorientation, 
in an approximation algorithm point of view~\footnote{\label{fn:approximation_alg}\url{https://en.wikipedia.org/wiki/Approximation_algorithm}}.

Intuitively, if the transform matrix is close to the identity matrix, then the reoriented scheme are closed to the original scheme which has been well optimized. 
This intuition can be quantitatively described in Proposition~\ref{prop:grad_reorientaiton}, 
which shows that if the transform matrix is inside a ball of the identity matrix with radius $\delta$, 
then the inner product of two reoriented gradient vectors $\Vx_1^T\Vx_2$ is inside a range centered at the inner product of the original two gradient vectors $\uu_1^T\uu_2$ with a radius of $2\delta+\delta^2$. 
The proof of this proposition is shown in Appendix~\ref{app:proof}.

\begin{proposition}{\textbf{Angular separation after gradient reorientation.}}\label{prop:grad_reorientaiton}
  Let  $\uu_1$ and $\uu_2$ be two gradient vectors in $\mathbb{S}^2$. 
  Denote the two gradient vectors after reorientation as $\Vx_1$ and $\Vx_2$, 
  and the corresponding transform matrices as $\BM{T}_1$ and $\BM{T_2}$ such that $\Vx_i=\BM{T}_i\uu_i$, $i=1,2$. 
  If $\|\BM{T}_i-\BM{I}\|_2\leq \delta$, where $\BM{I}$ is the identity matrix, 
  then the reoriented gradient vectors satisfies
  $ | \Vx_1^T\Vx_2 - \uu_1^T\uu_2 | \leq  2\delta +\delta^2$.  
\end{proposition}


Note that Proposition~\ref{prop:grad_reorientaiton} works for gradient reorientation using rotation matrices and also general affine transforms~\cite{Alexander_TMI2001}, 
where if $\BM{A}_i$ is the affine matrix used to reorient $\uu_i$ such that $\Vx_i=\frac{\BM{A}_i\uu_i}{\|\BM{A}_i\uu_i\|_2}$, then $\BM{T}_i=\frac{\BM{A}_i}{\|\BM{A}_i\uu_i\|_2}$. 
Proposition~\ref{prop:grad_reorientaiton} means that the separation angle of two reoriented gradients $\arccos(\Vx_1^T\Vx_2)$ is in $[\arccos(\uu_1^T\uu_2+2\delta+\delta^2),\arccos(\uu_1^T\uu_2-2\delta-\delta^2)]$. 
And the minimal separation angle of any two reoriented gradients is also in a range defined by the separation angle of two original gradient vectors. 
Thus, in an approximation algorithm point of view~\textsuperscript{\ref{fn:approximation_alg}}, maximizing the angular separation of the original scheme for scanner is to approximately maximize the angular separation of the reorientated scheme for model fitting. 
In practice, the transform matrix between a reoriented gradient vector and an original gradient vector in motion correction or eddy current correction is close to the identity matrix, 
i.e., $\|\BM{T}-\BM{I}\|_2<\delta$, and $\delta$ is a small value. 
If the transform matrix $\BM{T}$ is the rotation about $z$-axis with the degree of $5^\circ$, then $\epsilon=0.123$ that is about $12\%$ of the norm of $\BM{I}$. 
In practice, the rotation angle of the transform matrix is smaller than $5^\circ$. 
Thus the change of the separation angle after gradient reorientation is small. 

For the EEM formulation, as shown in~\EEqref{eq:EEM}, the electrostatic energy between $\uu_i$ and $\uu_j$ is 
\begin{footnotesize}
\begin{align}
  f_2(\uu_i,\uu_j) & =\frac{1}{\|\uu_i-\uu_j\|_2^2} + \frac{1}{\|\uu_i+\uu_j\|_2^2} = \frac{1}{1-(\uu_i^T\uu_j)^2} \nonumber \\
   & \in \left [\frac{1}{1- (|\uu_i^T\uu_j|-2\delta-\delta^2)^2}, \frac{1}{1- (|\uu_i^T\uu_j|+2\delta+\delta^2)^2}\right]
\end{align}
\end{footnotesize}
Based on Proposition~\ref{prop:grad_reorientaiton}, after gradient reorientation, the electrostatic energy between $\Vx_i$ and $\Vx_j$ is
\begin{footnotesize}
\begin{equation}
  f_2(\Vx_i,\Vx_j) = \frac{1}{1-(\Vx_i^T\Vx_j)^2}  \in \left [\frac{1}{1- (|\uu_i^T\uu_j|-2\delta-\delta^2)^2}, \frac{1}{1- (|\uu_i^T\uu_j|+2\delta+\delta^2)^2}\right]
\end{equation}
\end{footnotesize}%
Note that here we assume $|\uu_i^T\uu_j| + 2\delta+\delta^2<1$. 
The length of the range is 
\begin{footnotesize}
\begin{align}
  & \frac{1}{1- (|\uu_i^T\uu_j|+2\delta+\delta^2)^2} - \frac{1}{1- (|\uu_i^T\uu_j|-2\delta-\delta^2)^2} \nonumber \\ 
  &  =  \frac{4 |\uu_i^T\uu_j|(2\delta+\delta^2)}{ (1-(|\uu_i^T\uu_j|+2\delta+\delta^2)^2)(1-(|\uu_i^T\uu_j|-2\delta-\delta^2)^2) } \nonumber \\ 
  & \leq \frac{4c_0 (2\delta+\delta^2)}{(1-(c_0+2\delta+\delta^2)^2)(1-(c_0-2\delta-\delta^2)^2)} = L(c_0, \delta) \label{eq:seg_length}
\end{align}
\end{footnotesize}%
where $c_0=\max_{i< j} |\uu_i^T\uu_j|<1-2\delta-\delta^2$. 
Thus, the electrostatic energy of the reoriented sampling scheme $\sum_{i< j} f_2(\uu_i,\uu_j)$ , and the electrostatic energy of the original scheme $\sum_{i< j} f_2(\Vx_i,\Vx_j)$, 
are both inside a range, and we have 
\begin{equation}\label{eq:seg_length_total}
  \left |\sum_{i< j} f_2(\uu_i,\uu_j) - \sum_{i< j} f_2(\Vx_i,\Vx_j) \right | \leq K(K-1)L(c_0, \delta)
\end{equation}
where $K$ is the number of samples in the single shell scheme. 
Thus, minimizing the electrostatic energy of the original scheme is to approximately minimize the energy of the reoriented scheme. 
When $\delta$ is small, i.e., the transform matrix is close to the identity matrix, the range length in~\EEqref{eq:seg_length_total} is small and the electrostatic energies of the original scheme and the reoriented scheme are close.

We have shown that uniform sampling scheme design before gradient reorientation is an approximation of uniform sampling scheme design after gradient reorientation. 
However, it should be noted that if the transform matrix due to gradient reorientation is far from the identity matrix, the approximation error can be large. 
As shown in experiments, the proposed SC method yields larger angular separation (i.e., covering radius) between samples than the state-of-the-art EEM and GEEM method. 
If we assume that gradient reorientation reduces the covering radius of all schemes by the same amount ($<5^\circ$), then, after gradient reorientation, the scheme by SC still has larger covering radius than the scheme by EEM.

\subsubsection{Minimizing gradient heating}
The SC formulation in this paper is to design uniform sampling scheme in shells, 
  similarly with the works based on EEM~\cite{jones_MRM99,caruyer_CDMRI11,caruyer_MRM13,zhan:ISBI2011,koay_MP2012,koay_MRM2014,dubois:2006,cook_JMRI2007}, 
  which have not yet considered gradient heating specifically. 
  We have proposed to use ISC for P-O, where the first partial set of samples are nearly uniformly separated. 
  It was reported in~\cite{pierpaoli:2010:artifacts} that 
  optimal ordering not only ensures that usable data are available if the scan needs to be cut short, but also reduces the risk of systematic errors due to gradient heating and subject motion. 
  Minimizing gradient heating is beyond of the content of this paper, and gradient heating may be reduced in different ways~\cite{sjolund:JMR2015,hutter:ISMRM2017}. 
  It can be a future work to investigate which way of reducing gradient heating is more appropriate to be incorporated into current SC framework in this paper.

\section{Conclusion}
\label{sec:conclusion}

In this paper, we proposed to use the Spherical Code concept, which directly maximizes angular separations between samples, to design single- and multi-shell uniform sampling schemes.
We showed that various continuous and discrete sampling scheme design problems can be formulated using SC. 
We proposed five novel algorithms (i.e., ISC, IMOC, 1-Opt, MILP, CNLO) based on SC for uniform sampling scheme design. 
See Table~\ref{tab:algorithm_problems}. 
ISC is mainly to order samples in an existing scheme. 
IMOC and 1-Opt are to quickly obtain schemes in deterministic greedy ways. 
Although IMOC is a greedy incremental method, it obtains globally optimal solution in $\mathbb{S}^1$. 
For the sphere $\mathbb{S}^2$ in dMRI, covering radii of schemes by IMOC are very close to the best known solutions in single shell case~\textsuperscript{\ref{fn:neil}}, and are much larger than EEM in CAMINO. 
CNLO is a local optimization method which refines a given initialization (normally from IMOC+1-Opt). 
%
%
MILP is used to perform sub-sampling in an existing scheme.
It can obtain globally optimal solution or a good result within 10 minutes. 
The experiments showed that compared with the state-of-the-art methods~\cite{caruyer_MRM13,koay_MP2012} based on electrostatic energy which is used in HCP, 
the multi-shell scheme by IMOC + 1-Opt + CNLO yields larger covering radii in general, 
and better rotational invariance in dMRI reconstruction when the SNR is high and the number of samples is small. 
However, when a large number of noisy samples are used in DWI reconstruction, the sampling schemes by different methods may have similar reconstruction results. 
The proposed SC methods are to maximize the angular separation between samples, independent of diffusion models and diffusion reconstruction. 
The optimal sampling scheme that yields the best reconstruction quality is dependent on the diffusion model, reconstruction algorithm, and the ground truth diffusion signal. 
It is a future work to optimize the sampling schemes for a given specific diffusion model and a given reconstruction algorithm, so that the final result of the diffusion reconstruction is optimized. 
We have released some demos and codes in DMRITool~\textsuperscript{\ref{fn:dmritool}}. 

\appendices
\section{Proof of Proposition~\ref{prop:grad_reorientaiton}}
\label{app:proof}

Denote $\BM{T}_i=\BM{I}+\BM{C}_i$, then we have $\|\BM{C}_i\|_2\leq \delta$, and 
\begin{align}
  \Vx_1^T\Vx_2 &= \uu_1^T(\BM{I}+\BM{C}_1^T)(\BM{I}+\BM{C}_2)\uu_2 \nonumber \\
   & = \uu_1^T\uu_2 + \uu_1^T(\BM{C}_1^T+\BM{C}_2)\uu_2 + \uu_1^T\BM{C}_1^T\BM{C}_2\uu_2 \nonumber 
\end{align}
Let $c=|\uu_1^T(\BM{C}_1^T+\BM{C}_2)\uu_2| + |\uu_1^T\BM{C}_1^T\BM{C}_2\uu_2|$, then 
\begin{align*}
   | \Vx_1^T\Vx_2 - \uu_1^T\uu_2 | \leq c
\end{align*}
Because $|\uu_1^T(\BM{C}_1^T+\BM{C}_2)\uu_2|\leq \|\BM{C}_1^T+\BM{C}_2\|_2 \leq 2\delta$, 
  and $|\uu_1^T\BM{C}_1^T\BM{C}_2\uu_2| \leq \|\BM{C}_1^T\|_2\|\BM{C}_2\|_2=\delta^2$, we have $c\leq 2\delta+\delta^2$. 
  Then 
\begin{equation}
   | \Vx_1^T\Vx_2 - \uu_1^T\uu_2 |  \leq 2\delta +\delta^2
\end{equation}




\section*{Acknowledgments}
The authors would like to thank Dr.~Ruiliang Bai and Dr.~Alexandru Avram for useful discussions. 
This work was supported in part by funds provided by the Intramural Research Program of the \emph{Eunice Kennedy Shriver} National Institute of  Child  Health  and  Human  Development  (NICHD)  (ZIA-HD000266), 
and NIH grants (NS093842 and EB022880). 
The data were provided in part by the Human Connectome Project, WU-Minn Consortium 
(Principal Investigators: David Van Essen and Kamil Ugurbil; 1U54MH091657) funded by the 16 NIH Institutes and Centers that support the NIH Blueprint for Neuroscience Research; 
and by the McDonnell Center for Systems Neuroscience at Washington University.


%





\ifCLASSOPTIONcaptionsoff
  \newpage
\fi






\end{document}